\pdfoutput=1

\def\doi#1{doi:#1}
\def\qut#1{``#1''}

\documentclass[cp, online, hvmath]{copernicus}

\usepackage{pifont}

\begin{document}\sloppy

\title{Clouds and the Faint Young Sun Paradox}

\author[1,*]{C.~Goldblatt}
\author[1]{K.~J.~Zahnle}

\affil[1]{Space Science and Astrobiology Division, NASA Ames Research Center, MS
245-3,\newline Moffett Field, CA 94035, USA}
\affil[*]{now at: Astronomy Department, University of Washington, Box 351580, Seattle, WA 98195, USA}

\correspondence{C.~Goldblatt (cgoldbla@uw.edu)}

\runningtitle{Clouds and the Faint Young Sun Paradox}

\runningauthor{C.~Goldblatt and K.~J.~Zahnle}

\received{11 May 2010}
\accepted{25 May 2010}
\published{}

\firstpage{1}

\maketitle

\begin{abstract}
      We investigate the role which clouds could play in resolving the Faint
      Young Sun Paradox (FYSP). Lower solar luminosity in the past means
      that less energy was absorbed on Earth (a~forcing of
      ${-}$50\,{W\,m$^{-2}$} during the late Archean), but geological
      evidence points to the Earth being at least as warm as it is today,
      with only very occasional glaciations. We perform radiative
      calculations on a~single global mean atmospheric column. We select
      a~nominal set of three layered, randomly overlapping clouds, which are
      both consistent with observed cloud climatologies and reproduce the
      observed global mean energy budget of Earth. By varying the fraction,
      thickness, height and particle size of these clouds we conduct a~wide
      exploration of how changed clouds could affect climate, thus
      constraining how clouds could contribute to resolving the FYSP. Low
      clouds reflect sunlight but have little greenhouse effect. Removing
      them entirely gives a~forcing of ${+}$25\,{W\,m$^{-2}$} whilst more
      modest reduction in their efficacy gives a~forcing of ${+}$10 to
      ${+}$15\,{W\,m$^{-2}$}. For high clouds, the greenhouse effect
      dominates. It is possible to generate ${+}$50\,{W\,m$^{-2}$} forcing
      from enhancing these, but this requires making them 3.5 times thicker
      and 14\,K colder than the standard high cloud in our nominal set and
      expanding their coverage to 100\% of the sky. Such changes are not
      credible. More plausible changes would generate no more that
      ${+}$15\,{W\,m$^{-2}$} forcing. Thus neither fewer low clouds nor more
      high clouds can provide enough forcing to resolve the FYSP. Decreased
      surface albedo can contribute no more than ${+}$5\,{W\,m$^{-2}$}
      forcing. Some models which have been applied to the FYSP do not
      include clouds at all. These overestimate the forcing due to increased
      \chem{CO_2} by 20 to 25\% when $p\chem{CO_2}$ is 0.01 to 0.1\,bar.
\end{abstract}

\vspace*{-2.5mm}

\introduction\vspace*{-.5mm} Earth received considerably less energy from the
Sun early in Earth's history than today; ca.~2.5\,Ga (billion years before present) the sun was only 80\% as
bright as today. Yet the geological evidence suggests generally warm
conditions with only occasional glaciation. This apparent contradiction is known
as
      the Faint Young Sun Paradox \citep[FYSP,][]{ringwood-61,sm-72}. A~warm
      or temperate climate under a~faint sun implies that Earth had either
      a~stronger greenhouse effect or a~lower planetary albedo in the
      past, or both. In this study, we focus on the role of clouds in the FYSP. We
      both examine how their representation in models affects calculations
      of changes in the greenhouse effect, and constrain the direct
      contribution that changing clouds could make to resolving the FYSP.

      Clouds have two contrasting radiative effects. In the spectral region
      of solar radiation (shortwave hereafter) clouds are highly
      reflective. Hence clouds contribute a~large part of Earth's planetary
      albedo (specifically the Bond albedo, which refers to the fraction of
      incident sunlight of all wavelengths reflected by the planet). In the
      spectral region of terrestrial thermal radiation (longwave hereafter),
      clouds are a~strong radiative absorber, contributing significantly to
      the greenhouse effect. Cloud absorption is largely independent of
      wavelength (they approximate to \qut{grey} absorbers), in contrast to
      gaseous absorbers which absorb only in certain spectral regions
      corresponding to the vibration--rotation lines of the molecules.

      Despite the obvious, first-order, importance of clouds in climate, it
      has become conventional to omit them in models of early Earth climate
      and use instead an artificially high surface albedo. As described by
      \citet{kpc-84}:
\begin{quote}
\textit{Clouds are not included explicitly in the model; however, their
effect on the radiation balance is accounted for by adjusting the effective
albedo to yield a~mean surface temperature of 288\,K for the present Earth.
The albedo is then held fixed for all calculations at reduced solar fluxes.
$\ldots$we feel that the assumption of constant albedo is as good as can be
done, given the large uncertainties in the effect of cloud and ice albedo
feedbacks.}
\end{quote}
In effect, the surface is whitewashed in lieu of putting clouds in the
atmosphere. This assumption has been used extensively in the models from Jim
Kasting's group
(\citealp{kpc-84,ka-86,k-87,k-88,kwr-93,pav-ea-00,phka-03,kh-06,hdkk-08}),
which, together with parametrisations and results based on these models (for
example,
\citealp{ktp-88,ck-92a,ck-92b,kps-01,k-05,tajika-03,bb-02,l-00,franck-ea-99,franck-ea-00,vb-ea2-03,vb-ea-03,lvb-01,blw-04})
have dominated early Earth palaeoclimate and other long term climate change
research for the last two and a~half decades. The validity of this method has
not previously been tested.

      Whilst Kasting's approach is that changes to clouds are so difficult
      to constrain that one cannot justifiably invoke them to resolve the
      FYSP, others are more bold. Some recent papers have proposed
      cloud-based resolutions to the FYSP.

      \citet{ron-lindzen-10} focus on increasing the warming effect of high
      clouds, finding that a~total covering of high clouds which have been
      optimised for their warming effect could give a~late Archean global
      mean temperature at freezing without increasing greenhouse gases. To
      justify such extensive clouds, they invoke the \qut{iris} hypothesis
      \citep{lindzen-ea-01} which postulates that cirrus coverage should increase if surface
      temperatures decrease \citep[this hypothesis has received much criticism,
      e.g.][]{hartmann-michelsen-02,chambers-ea-02}.

      \citet{rosing-ea-10} focus on decreasing the reflectivity of low level
      clouds so that the Earth absorbs more solar radiation. To justify
      this, they suggest that there was no emission of the important
      biogenic cloud condensation nuclei (CCN) precursor dimethyl sulphide
      (DMS) during the Archean and, consequently, clouds were thinner and
      had larger particle sizes.

      We note that both \citet{ron-lindzen-10} and \citet{rosing-ea-10}
      predict early Earth temperatures substantially below today's, which we
      do not consider a~satisfactory resolution of the FYSP.

\citet{shaviv-03} and \citet{svensmark-07} proposes less low-level cloud on early Earth due to fewer galactic cosmic rays being incident on the lower troposphere. The underlying hypothesis is of a correlation between galactic cosmic ray incidence and stratus amount, through CCN creation due to tropospheric ionization \citep{svensmark-friis-97,svensmark-07}. This has received extensive study in relation to contemporary climate change and has been refuted \citep[e.g.][and references therein]{sun-bradley-02,lockwood-frohlich-07,kris-ea-08,bailerjones-09,calogovic-ea-10,kulmala-ea-10}.

      In this study, we comprehensively asses how the radiative properties
      of clouds, and changes to these, can affect the FYSP.  First, we
      explicitly evaluate how accurate cloud-free calculations of changes in
      the greenhouse effect are with respect to atmospheres with clouds
      included. We do this by considering a~very wide range of cloud
      properties within a~single global mean atmospheric column, finding
      a~case study which matches Earth's energy budget, then comparing the
      effect of more greenhouse gas in this column to a~cloud-free
      calculation. Second, we conduct a~very wide exploration of how
      changing clouds could directly influence climate. We vary fraction,
      thickness, height and particle size of the clouds and vary surface
      albedo. We do not advocate any particular set of changes to
      clouds. Rather, constraints on what direct contribution clouds can make towards
      resolving the FYSP emerges from our wide exploration of the
      phase space.

The heyday of clouds in 1-D models was the 1970s and 1980s. Improvements in radiative transfer codes and computational power over the last 30 years allow us to contribute new insight to the problem, in particular by widely exploring phase-space. Nonetheless, these classic papers retain their relevence and are instructive as to how one might treat clouds in such simple models \citep[e.g.][]{schneider-72,reck-79,wang-stone-80,stephens-webster-81,charlock-82}. With specific relevance to the FYSP, \citet{kd-87} included clouds in their model of methane and carbon dioxide warming on early Earth, and calculated radiative forcings from some changed cloud cases (our results here agree with this older work). \citet{rhsw-82} considered cloud feedbacks for early Earth. 

      Regarding whether a~cloud-free model will correctly calculate the
      increased greenhouse effect with increases gaseous absorbers, we
      hypothesise that it will lead to an overestimation in the efficacy of
      enhanced greenhouse gases.  In the absence of clouds, the broadest
      range of absorption is due to water vapour. However, whilst water
      vapour absorbs strongly at shorter and longer wavelengths, it absorbs
      weakly between 8 and 15\,{\unit{\mu}m}. This region of weak absorption is known
      as the water vapour window. It is coincident with the Wein peak of
      Earth's surface thermal emission at 10\,{\unit{\mu}m}. Thus the water vapour
      window permits a~great deal of surface radiation to escape to space
      unhindered. Other greenhouse gases -- and clouds -- do absorb here, so
      are especially important to the greenhouse effect. With clouds
      absorbing some fraction of the radiation at all wavelengths, the
      increase in absorption with increased greenhouse gas concentration
      would be less than if clouds were absent. Therefore, we think that
      a~cloud-free model would overestimate increased gaseous absorption
      with increased greenhouse gas abundance and underestimate the
      greenhouse gas concentrations required to keep early Earth warm.

      Comparison of cloudy and cloud-free radiative forcings in the context
      of anthropogenic climate change
      \citep{pinnock-ea-95,myhre-stordal-97,jbmw-00} supports our
      hypothesis. For \chem{CO_2}, a~clear-sky calculation overestimates the
      radiative forcing by 14\%. For more exotic greenhouse gases, which are
      optically thin at standard conditions (CFCs, CCs, HCFCs, HFCs, PFCs,
      bromocarbons, iodocarbons), clear-sky calculations overestimate
      radiative forcing by 26--35\%. \chem{CH_4} and \chem{N_2O} are
      intermediate; their clear-sky radiative forcings are overestimated by
      29\% and 25\%, respectively \citep{jbmw-00}.

      A~roadmap of our paper is as follows. In Sect.~\ref{s-methods} we
      describe our general methods, verification of the radiative transfer
      scheme and the atmospheric profile we use. In Sect.~\ref{s-cloudrep}
      we deal specifically with the development of a~case study of three
      cloud layers representing the present climate and the model
      sensitivity to this. In Sect.~\ref{s-kastingeval} we compare cloudy
      and cloud-free calculations of the forcing from increased greenhouse
      gas concentration. In Sect.~\ref{s-cloudpropvar} we explore what
      direct forcing clouds could impart, and in section
      \ref{s-fys_resol_eval} we evaluate the aforementioned cloud-based
      hypotheses for resolving the FYSP.

\section{Methods}
\label{s-methods}

\subsection{Overview}

      Using a~freely available radiative transfer code, we develop a~set of
      cloud profiles for single-column models which is in agreement both
      with cloud climatology and the global mean energy budget. This serves
      as the basis for comparison of radiative forcing with a~clear sky
      model and with changed cloud properties.

\subsection{Radiative forcing}

      In work on contemporary climatic change, extensive use is made of
      \textit{radiative forcing} to compare the efficacy of greenhouse gases
      \citep[e.g.][]{ar4-2}. This is defined as the change in the net flux
      at the tropopause with a~change in greenhouse gas concentration,
      calculated either on a~single fixed temperature--pressure profile or
      a~set of fixed profiles and in the absence of climate
      feedbacks. Surface temperature change is directly proportional to
      radiative forcing, with a~radiative forcing of approximately
      5\,{W\,m$^{-2}$} being required to cause a~surface temperature change of
      1\,K \citep[see Fig.~7 of][]{glw-09}. Note that the tropopause must
      be defined as the level at which radiative heating becomes the
      dominant diabatic heating term \citep{ffs-97}, i.e.\ the lowest level
      at which the atmosphere is in radiative equilibrium.

      We base all our analyses on radiative forcings here. As we make
      millions of radiative transfer code evaluations, saving in
      computational cost from comparing radiative forcings rather than
      running a~radiative-convective climate model is significant, and
      facilitates the wide range of comparisons presented.

\subsection{Global Annual Mean atmosphere}

\def\yx{\kern5.5pt}
\def\dt{\kern1.5pt}
\begin{table}[t]
\caption{GAM profile at levels (layer boundaries). Note that the tropopause is at 100\,hPa.} \label{t-GAMlevels}
\vskip4mm
\begin{center}
\scalebox{.90}[.90]{%
\begin{tabular}{rrrrr}
\tophline
&&&&\\[-12pt]
\multicolumn{1}{c}{Pressure} & \multicolumn{1}{c}{Altitude} & \multicolumn{1}{c}{Temperature} & \multicolumn{1}{c}{Water vapour} & \multicolumn{1}{c}{Ozone} \\
\multicolumn{1}{c}{(Pa)} & \multicolumn{1}{c}{(km)} & \multicolumn{1}{c}{(K)} & \multicolumn{1}{c}{(g/kg)} & \multicolumn{1}{c}{(ppmv)} \\
\middlehline
    10{\yx}{\yx} & 64.739{\yx}{\yx} & 230.00{\yx}{\yx} &  0.0036{\yx}{\yx} & 1.080{\yx}{\yx}{\yx}\\
    20{\yx}{\yx} & 59.912{\yx}{\yx} & 245.61{\yx}{\yx} &  0.0036{\yx}{\yx} & 1.384{\yx}{\yx}{\yx}\\
    30{\yx}{\yx} & 56.951{\yx}{\yx} & 252.88{\yx}{\yx} &  0.0036{\yx}{\yx} & 1.626{\yx}{\yx}{\yx}\\
    50{\yx}{\yx} & 53.114{\yx}{\yx} & 260.00{\yx}{\yx} &  0.0036{\yx}{\yx} & 1.974{\yx}{\yx}{\yx}\\
   100{\yx}{\yx} & 47.763{\yx}{\yx} & 266.29{\yx}{\yx} &  0.0035{\yx}{\yx} & 2.600{\yx}{\yx}{\yx}\\
   200{\yx}{\yx} & 42.393{\yx}{\yx} & 260.33{\yx}{\yx} &  0.0033{\yx}{\yx} & 5.484{\yx}{\yx}{\yx}\\
   300{\yx}{\yx} & 39.339{\yx}{\yx} & 254.31{\yx}{\yx} &  0.0032{\yx}{\yx} & 6.810{\yx}{\yx}{\yx}\\
   500{\yx}{\yx} & 35.612{\yx}{\yx} & 243.24{\yx}{\yx} &  0.0032{\yx}{\yx} & 7.242{\yx}{\yx}{\yx}\\
  1000{\yx}{\yx} & 30.842{\yx}{\yx} & 228.11{\yx}{\yx} &  0.0031{\yx}{\yx} & 7.490{\yx}{\yx}{\yx}\\
  2000{\yx}{\yx} & 26.290{\yx}{\yx} & 222.10{\yx}{\yx} &  0.0030{\yx}{\yx} & 6.169{\yx}{\yx}{\yx}\\
  3000{\yx}{\yx} & 23.671{\yx}{\yx} & 218.71{\yx}{\yx} &  0.0029{\yx}{\yx} & 4.780{\yx}{\yx}{\yx}\\
  5000{\yx}{\yx} & 20.445{\yx}{\yx} & 212.59{\yx}{\yx} &  0.0026{\yx}{\yx} & 2.250{\yx}{\yx}{\yx}\\
 10000{\yx}{\yx} & 16.204{\yx}{\yx} & 206.89{\yx}{\yx} &  0.0023{\yx}{\yx} & 0.516{\yx}{\yx}{\yx}\\
 15000{\yx}{\yx} & 13.727{\yx}{\yx} & 211.83{\yx}{\yx} &  0.0048{\yx}{\yx} & 0.344{\yx}{\yx}{\yx}\\
 20000{\yx}{\yx} & 11.914{\yx}{\yx} & 219.01{\yx}{\yx} &  0.0153{\yx}{\yx} & 0.160{\yx}{\yx}{\yx}\\
 25000{\yx}{\yx} & 10.461{\yx}{\yx} & 225.87{\yx}{\yx} &  0.0456{\yx}{\yx} & 0.122{\yx}{\yx}{\yx}\\
 30000{\yx}{\yx} &  9.237{\yx}{\yx} & 233.27{\yx}{\yx} &  0.1852{\yx}{\yx} & 0.089{\yx}{\yx}{\yx}\\
 35000{\yx}{\yx} &  8.168{\yx}{\yx} & 240.52{\yx}{\yx} &  0.3751{\yx}{\yx} & 0.070{\yx}{\yx}{\yx}\\
 40000{\yx}{\yx} &  7.215{\yx}{\yx} & 247.19{\yx}{\yx} &  0.6046{\yx}{\yx} & 0.058{\yx}{\yx}{\yx}\\
 45000{\yx}{\yx} &  6.352{\yx}{\yx} & 253.27{\yx}{\yx} &  0.8866{\yx}{\yx} & 0.051{\yx}{\yx}{\yx}\\
 50000{\yx}{\yx} &  5.562{\yx}{\yx} & 258.62{\yx}{\yx} &  1.2365{\yx}{\yx} & 0.047{\yx}{\yx}{\yx}\\
 55000{\yx}{\yx} &  4.834{\yx}{\yx} & 263.15{\yx}{\yx} &  1.6525{\yx}{\yx} & 0.045{\yx}{\yx}{\yx}\\
 60000{\yx}{\yx} &  4.159{\yx}{\yx} & 267.14{\yx}{\yx} &  2.1423{\yx}{\yx} & 0.045{\yx}{\yx}{\yx}\\
 65000{\yx}{\yx} &  3.529{\yx}{\yx} & 270.73{\yx}{\yx} &  2.7049{\yx}{\yx} & 0.044{\yx}{\yx}{\yx}\\
 70000{\yx}{\yx} &  2.938{\yx}{\yx} & 274.00{\yx}{\yx} &  3.3366{\yx}{\yx} & 0.042{\yx}{\yx}{\yx}\\
 75000{\yx}{\yx} &  2.381{\yx}{\yx} & 277.05{\yx}{\yx} &  4.1602{\yx}{\yx} & 0.039{\yx}{\yx}{\yx}\\
 80000{\yx}{\yx} &  1.855{\yx}{\yx} & 279.84{\yx}{\yx} &  5.2152{\yx}{\yx} & 0.035{\yx}{\yx}{\yx}\\
 85000{\yx}{\yx} &  1.356{\yx}{\yx} & 282.28{\yx}{\yx} &  6.3997{\yx}{\yx} & 0.033{\yx}{\yx}{\yx}\\
 90000{\yx}{\yx} &  0.882{\yx}{\yx} & 284.08{\yx}{\yx} &  7.8771{\yx}{\yx} & 0.032{\yx}{\yx}{\yx}\\
 95000{\yx}{\yx} &  0.431{\yx}{\yx} & 285.85{\yx}{\yx} &  9.5702{\yx}{\yx} & 0.031{\yx}{\yx}{\yx}\\

&
\hphantom{1234567890}&
\hphantom{1234567890}&
\hphantom{1234567890}&
\hphantom{1234567890}\\[-12pt]

100000{\yx}{\yx} &  0.000{\yx}{\yx} & 289.00{\yx}{\yx} & 11.1811{\yx}{\yx} & 0.031{\yx}{\yx}{\yx}\\
\bottomhline
\end{tabular}
}
\end{center}
\end{table}

      We perform all our radiative transfer calculations on a~single Global
      Annual Mean (GAM) atmospheric profile (Table~\ref{t-GAMlevels}). This
      is based on the GAM profile of \citet{christidis-ea-97} with some
      additional high altitude data from \citet{jbmw-00}. Surface albedo is
      set as 0.125 \citep{tfk-09}. For standard conditions we use year 2000
      gas compositions: 369\,ppmv \chem{CO_2}, 1760\,ppbv \chem{CH_4} and,
      316\,ppbv \chem{N_2O}. We use present day oxygen and ozone
      compositions throughout the work. 
Whilst comparisons without these might be interesting, they would necessitate using a different temperature profile in order to be self consistent. This would significantly complicate our methods, so no such calculations are performed. 
      For solar calculations, we use the
      present solar flux and a~zenith angle of 60{\degree}.

      Calculating radiative forcings on a~single profile does introduce some
      error relative to using a~set of profiles for various latitudes
      \citep{myhre-stordal-97,frec-ea-98,jbmw-00}. However, as this is
      a~methodological paper concerning single column radiative-convective
      models, it is the appropriate approach to take here.

\subsection{Radiative transfer code and verification}

      We use the Atmosphere Environment Research (AER) Rapid Radiative
      Transfer Model \citep[RRTM,][]{mlawer-ea-97,clough-ea-05-aercodes},
      longwave version 3.0 and shortwave version 2.5, which are available
      from \url{http://rtweb.aer.com} (despite different version numbers,
      these were both the most recent versions at the time of the
      research). RRTM has been parameterised for pressures between 0.01 and
      1050\,hPa and for temperatures deviating no more than 30\,K from the
      standard mid-latitude summer (MLS) profile. We have verified that the
      GAM profile we use is within this region of pressure-temperature
      space. The cloud parameterisations in RRTM which we select follow
      \citet{hu-stamnes-93} for water clouds and \citet{fu-yang-sun-98} for
      ice clouds.

\begin{figure*}
\vspace*{2mm}
\center
\includegraphics{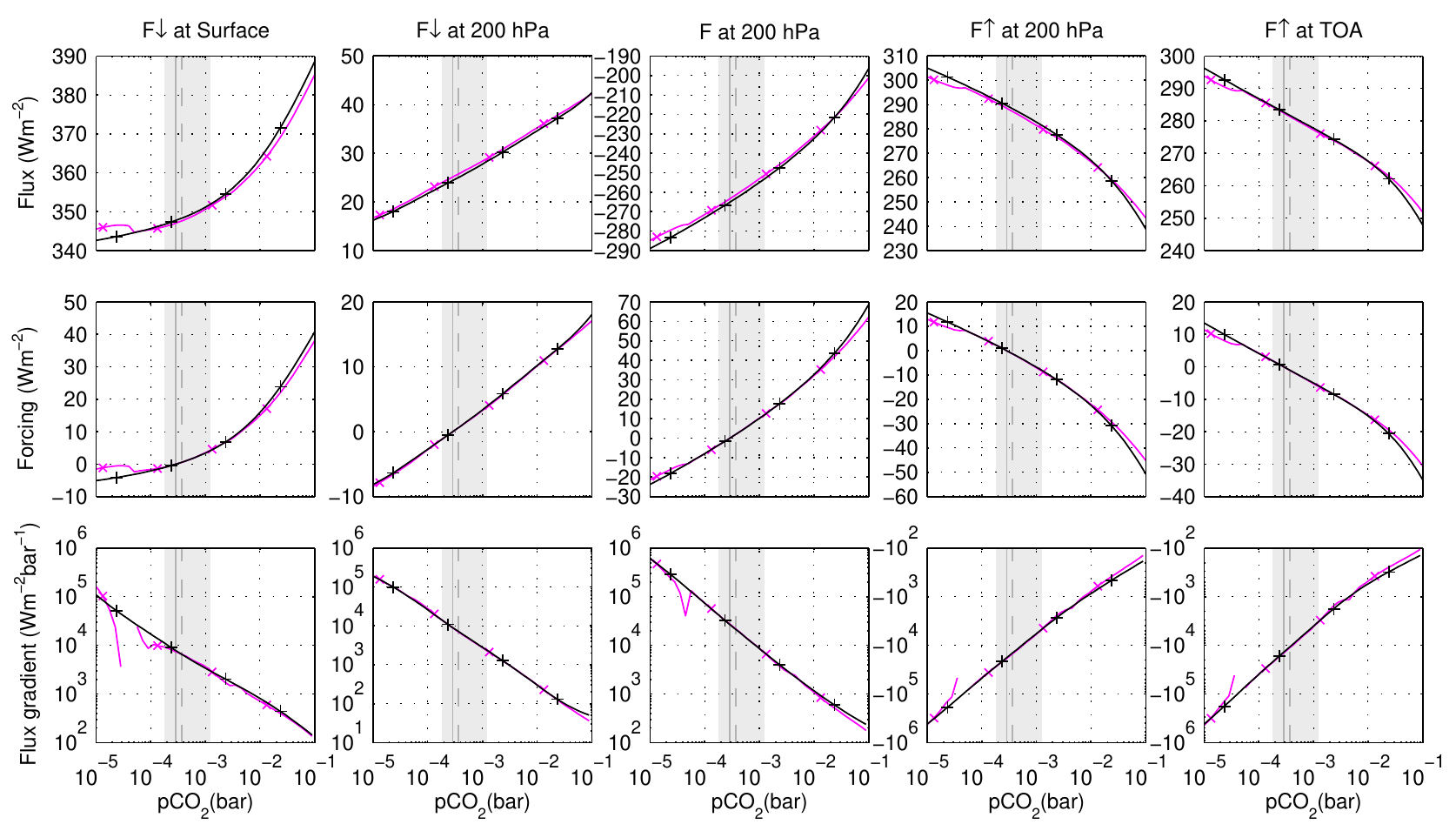}
\caption{RRTM performance for \chem{CO_2} for each flux and each level.
Colours (online only) and markers are: black ${+}$ for LBLRTM, magenta
${\times}$ for RRTM. Shaded areas are range from Quaternary minimum
(180\,ppmv) to SRES maximum (1248\,ppmv) concentration. Grey lines in these
areas are solid for pre-industrial (287\,ppmv) and dashed for year 2000
(369\,ppmv) concentrations.}
\label{f-verify-co2}
\end{figure*}
      RRTM has been designed primarily for contemporary atmospheric
      composition. Our intended use is for different atmospheric composition
      (higher greenhouse gas concentrations), so it is necessary for us to
      independently test the performance of the model under these conditions \citep{collins-ea-06, glw-09}. Following the
      approach of \citet{glw-09} we directly compare longwave clear sky
      radiative forcings from RRTM to the AER Line-by-Line Radiative
      Transfer Model \citep[LBLRTM,][]{clough-ea-05-aercodes}. These runs are
      done on a~standard Mid-Latitude Summer (MLS) profile
      \citep{mcclatchey-ea-71,ander-ea-86-stdatmdef} to take advantage of
      the large number of computationally expensive LBLRTM runs performed by
      \citet{glw-09}. Performance of the codes is evaluated at three levels:
      the top of the atmosphere (TOA), the MLS tropopause at 200\,hPa and
      the surface. Upward and downward fluxes are considered separately. The
      surface is taken to be a~black body, so the upward flux depends only
      on temperature ($F_\mathrm{lw,surf}^\uparrow = \sigma T_*^4$). The
      downward longwave flux at the TOA is zero. Neither vary with
      greenhouse gas concentrations, so changes in the net flux at these
      levels depends on one radiation stream only. At the tropopause the net
      flux is the sum of the two streams. It is defined positive downwards,
\begin{equation}
 F_{\rm lw} = F_{\rm lw}^\downarrow - F_{\rm lw}^\uparrow {} \, .
\end{equation}
      In addition to the radiative flux, we show (Fig.~\ref{f-verify-co2})
      the forcing
\begin{equation}
\mathbb{F}_{\rm lw} = F_{\rm lw} - F_{\rm lw,std} {}\, ,
\end{equation}
      where $F_{\rm lw,std}$ is the flux at preindustrial conditions and the
      flux gradient (change of flux with changing gas concentration)
\begin{equation}
\frac{\partial F}{\partial X} \approx  \frac{\Delta F}{\Delta X} =  \frac{F_{i+1} - F_{i}}{X_{i+1} - X_{i}} {}\, ,
\end{equation}
      where $F_i$ is the flux at gas concentration $X_i$ \citep{glw-09}.

      Our focus is on comparison of cloudy to cloud-free profiles within
      RRTM, so we do not need high accuracy calculations of early Earth
      radiative forcings. We can therefore use rather relaxed and
      qualitative thresholds for acceptable model performance relative to
      LBLRTM: we require continuous and monotonic response to changing
      greenhouse gas concentration (no saturation), the forcing should be
      smooth and monotonic and divergence from the LBLRTM flux gradient
      should be limited. For \chem{CO_2}, RRTM forcing is not smooth or
      monotonic below $p\chem{CO_2}{=}10^{-4}$\,bar so this region is
      excluded (see Fig.~\ref{f-verify-co2}). \chem{CO_2} concentrations up
      to $p\chem{CO_2}{=}10^{-1}$\,bar are used, though there is some
      underestimation of radiative forcing by RRTM above
      $p\chem{CO_2}{=}10^{-2}$\,bar. Also, collision-induced absorption (absorption due to forbidden transitions) becomes important at 
$p\chem{CO_2} \sim 0.1$\,bar 
\citetext{J. Kasting, private communication} but coefficients for these are
      not included in the HITRAN database on which both RRTM and LBLRTM
      absorption coefficients are based. Therefore, it is emphasised that the
      radiative forcings presented here for high \chem{CO_2} will be
      underestimates, but valid for intra-comparison.

      The comparison of RRTM to LBLRTM (Fig.~\ref{f-verify-co2}) is only for
      the purpose of validating clear sky radiative forcing in the context
      of this methodological study. We have undoubtedly used RRTM outside
      its design range. This is not intended as an assessment of its use for
      the contemporary atmosphere or for anthropogenic climate change.

\section{Cloud representation and model tuning}
\label{s-cloudrep}

\subsection{Practical problems and observational guidance}

      Generation of an appropriate cloud climatology for this work is not
      straightforward. Two fundamental problems are shortcomings in
      available cloud climatologies and averaging to a~single
      profile. Concerning climatologies, the problem is one of observations:
      surface observers will see the lowest level cloud only, satellites
      will see the highest level of cloud only. Radiosondes are cloud
      penetrating and cloud properties may be inferred from measured
      relative humidity, but the spatial and temporal coverage of radiosonde
      stations is limited. See \citet{wrz-00} and \citet{rzw-05} for
      extensive discussion of what progress can be made. Similarly, radar
      can profile clouds, but such observations are sparse. Concerning
      averaging, the dependence of the global energy budget on cloud
      properties is expected to be non-linear: one should not expect that
      a~linear average of global cloud physical properties would translate
      into a~set of clouds whose radiative properties would give energy
      balance in a~single column. Nonetheless, available temporally and
      spatially averaged data for cloud properties can guide how clouds should be
      represented in the model.

\begin{figure}
\vspace*{2mm}
\center
\includegraphics[width=8cm]{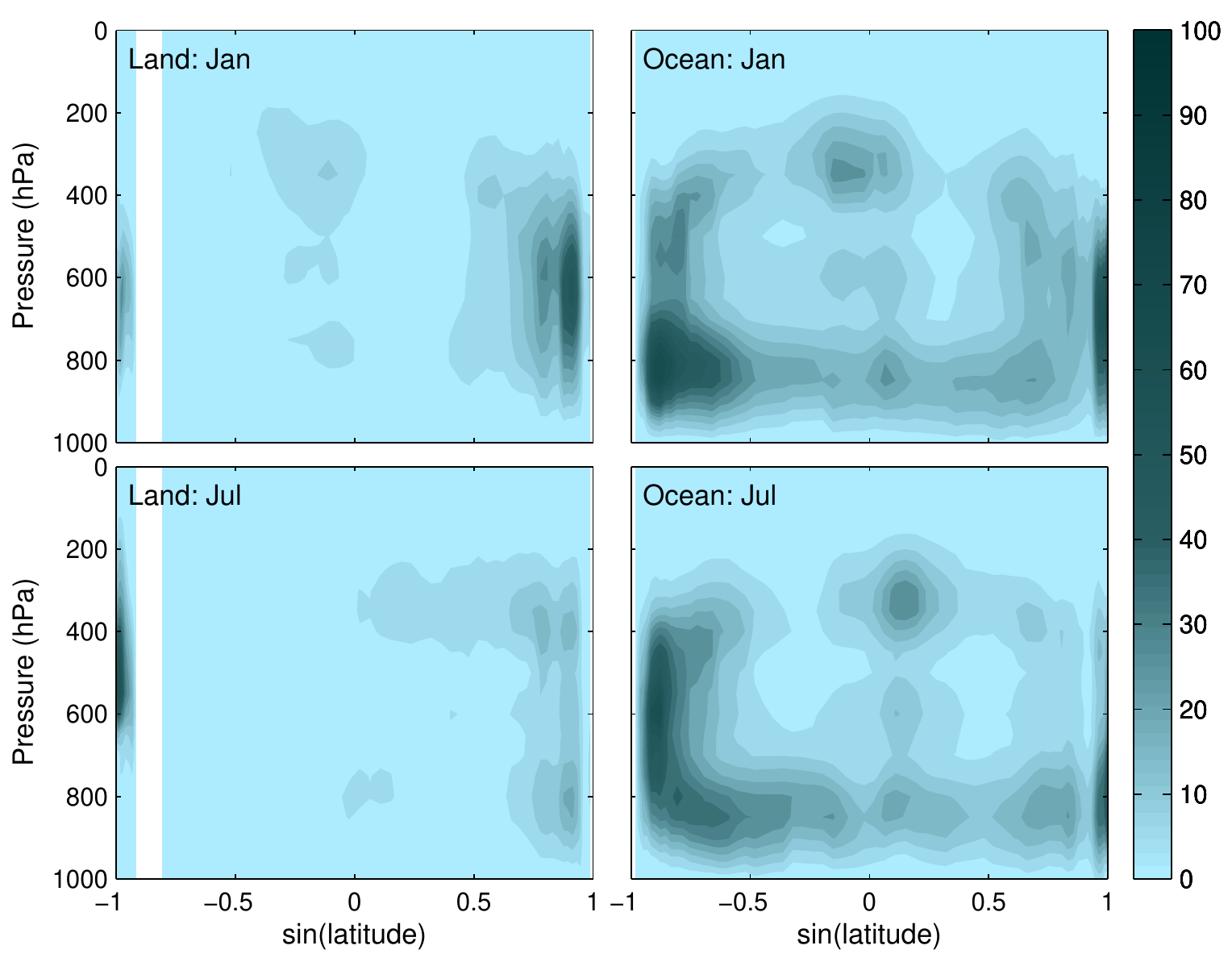}
\caption{Average cloud fraction with altitude following \citet[][and
W.~Rossow, personal communication, 2009]{rzw-05}, for January and July, land
and ocean. White areas are where there is either no land (the Southern and
Arctic Oceans) or no ocean (Antarctica).} \label{f-cloudfrac_4pcol}
\end{figure}
\begin{figure}
\vspace*{2mm}
\center
\includegraphics[width=8cm]{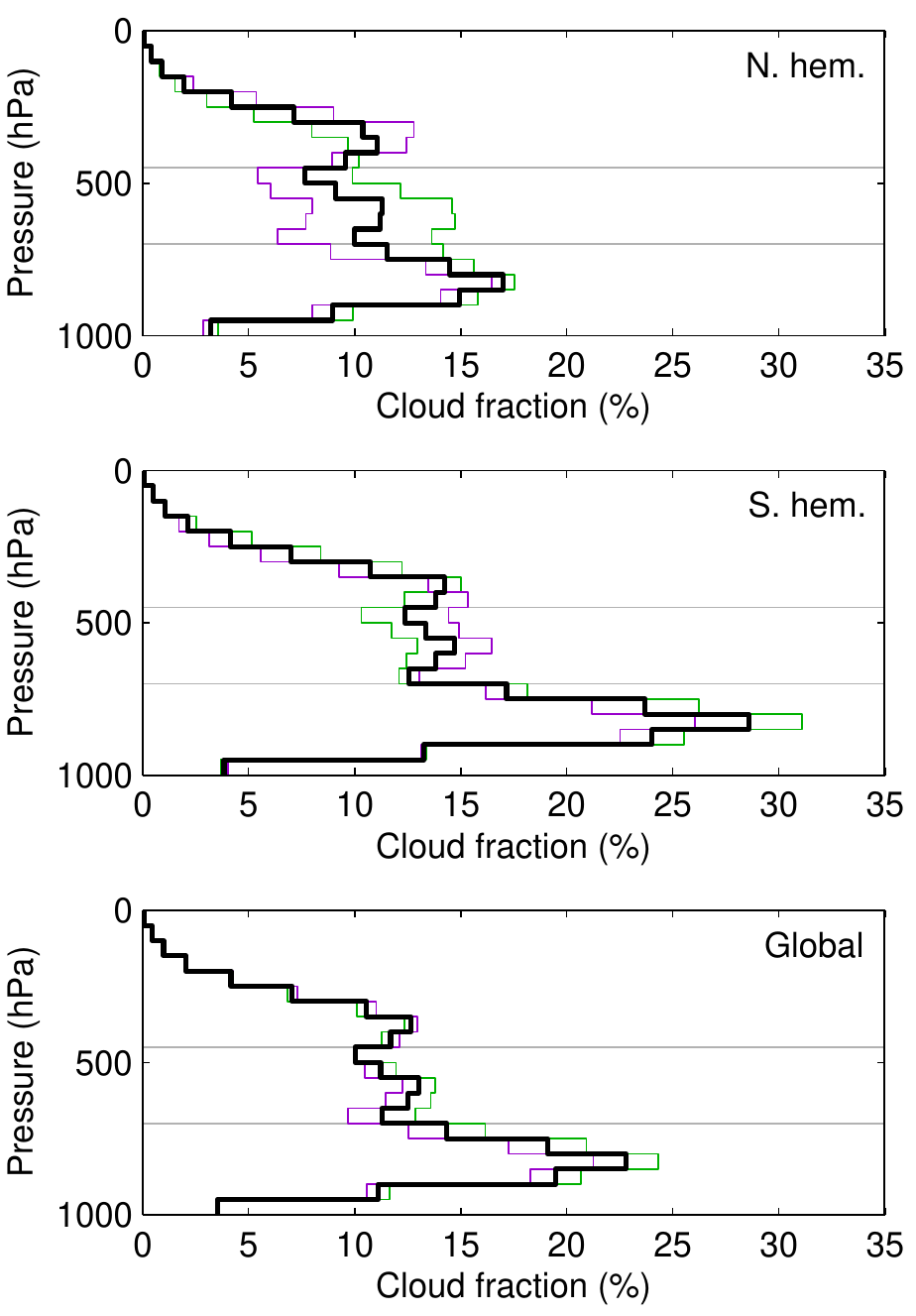}
\caption{Average cloud fraction with altitude following \citet[][and
W.~Rossow, personal communication, 2009]{rzw-05}. \textbf{(a)}~Northern
Hemisphere. \textbf{(b)}~Southern Hemisphere and \textbf{(c)}~global for
January (green), July (purple) and mean (black). Grey horizontal lines
separate high, mid- and low-level clouds.} \label{f-cloudfrac}
\end{figure}
      \citet{rzw-05} deduce zonally-averaged cloud fraction profiles using
      a~combination of International Satellite Cloud Climatology Program
      (ISCCP) and radiosonde data (Fig.~\ref{f-cloudfrac_4pcol}). The
      existence of three distinct cloud layers and the pressure levels of
      these are immediately apparent when averaging the data meridionally
      (Fig.~\ref{f-cloudfrac}). Following \citet{rs-99}, we divide the
      clouds into three groups, with divisions at 450 and
      700\,hPa. \qut{Low} cloud coresponds to cumulus, stratocumulus and
      stratus clouds. \qut{Mid} level clouds correspond to altocumulus,
      altostratus and nimbostratus. \qut{High} clouds correspond to cirrus
      and cirrostratus. Absolute cloud fractions cannot be extracted
      directly from these data as information on how the clouds overlap is
      lost in temporal and spatial averaging. A~simple approach to give
      indicative values is to assume either maximum or random overlap within
      each group (high, middle and low), then to scale these cloud amounts
      by a~constant such that randomly overlapping the three groups gives
      the IPCC mean global cloud fraction of 67.6\% \citep{rs-99}. Maximum
      and random overlap within groups give cloud fractions $[f_{\rm
      high},f_{\rm mid},f_{\rm low}]{=}[0.24,0.25,0.43]$ and $[f_{\rm
      high},f_{\rm mid},f_{\rm low}]{=}[0.25,0.29,0.39]$, respectively.

Averaged cloud optical thickness or water paths are more difficult to
constrain, as they are not directly available from the \citet{rzw-05} data
set \citetext{W.~Rossow, personal communication, 2009}. We proceed with ISCCP
data only. \citet{rs-99} report water paths of $[\, \mathcal{W}_{\rm high},
\mathcal{W}_{\rm mid}, \mathcal{W}_{\rm low}]{=}[23, 60 ,51]$\,{g\,m$^{-2}$}.
ISCCP data are from downward looking satellite data only and overlap is not
accounted for. Whilst the low cloud value will indicate low clouds only, high
and mid level cloud values may include opacity contributions from the lower
clouds which they obscure (see Fig.~\ref{f-cloudfrac_4pcol}). Hence these
water paths are indicative only.

      Using fine resolution spatially and temporally resolved data would
      help resolve these issues. However, to do so would be beyond the scope
      of this work and, we feel, it is beyond what is necessary to address
      the first-order questions which are the subject of this paper.

\vspace*{1.5mm}

\subsection{Development of cloud profiles}\vspace*{1mm}

      We need to develop a~set of cloud profiles which appropriately
      represents Earth's cloud and energy budget climatologies. By
      necessity, we shall need to simplify cloud properties, tune our model
      clouds and consider sensitivity of the model energy budget to these
      clouds.

Even with the assumption that each cloud is homogeneous, each of our three
cloud layers is represented by a~cloud base and top, water path, liquid:ice
ratio, and effective particle sizes for liquid and ice particles, giving
6\,degrees of freedom for each cloud. With three layers, there are eight
permutations for overlap, contributing another 7\,degrees of freedom for the
fractional coverage. A~total of 25\,degrees of freedom is clearly impossible
to explore fully. As a~necessary simplification, we fix the cloud base and
top, take clouds to be either liquid (low and mid clouds) or ice (high
clouds) and fix the particle size \mbox{\citep[following][]{rs-99}.} We
assume that cloud layers are randomly overlapped, so each cloud layer can be
represented by a~single fraction from which the overlap is calculated (many
GCMs use a \qut{maximum-random} overlap method where cloud \mbox{fractions in
adjacent} layers are correlated; this is not relevant here as our discrete
cloud layers are separated by intervals of clear sky, e.g.\ see
\citealp{hogan-illingworth-00}).

Random overlap is easiest to explain for the case of two cloud levels (A and
B), with cloud fractions $a$ and $b$. Fraction $ab$ of the sky would have
both cloud layers, fraction $a(1{-}b)$ would only have level~A clouds and
fraction $(1{-}a)b$ would only have level~B clouds, and fraction
$(1{-}a)(1{-}b)$ would be cloud free. With three cloud layers, we have eight
columns. Each column is evaluated separately in both longwave and shortwave
spectral regions and the final single column is found as a~weighted sum of
these 16\,evaluations. Different cloud fractions can be accounted for in this
summation, reducing the number of RRTM evaluations needed.

\def\yx{\kern5.5pt}
\def\dt{\kern1.5pt}
\begin{table}[t]
\caption{Parameter values used in the large cloud tuning ensemble. Optical
depth depends logarithmically on water path. Water path and cloud fraction
are varied independently for each layer. Water paths range from optically
thin to optically thick clouds \citep{cw-99} with 10 values. Cloud fractions
range from 5\% to 100\% coverage with 20 cases. Effective radius is for water
clouds (low and mid level) and generalised effective size is for ice clouds
(high). There there are $10^3 {\times} 20^3 {=} 8{\times}10^{6}$ cases in
total.} \label{t-largetune} \vskip4mm
\begin{center}
\begin{tabular}{lccc}
\tophline
Fixed properties & High & Mid & Low \\
\middlehline
Cloud top (hPa) & 300 & 550 & 750 \\
Cloud base (hPa) & 350 & 650 & 900 \\
Liquid or ice & Ice & Liquid & Liquid \\
Effective radius (\unit{\mu }m) & -- & {\yx}11 & {\yx}11 \\
Generalised effective size (\unit{\mu }m)  & {\yx}75 & -- & -- \\[1.5mm]
\middlehline
Variable properties & \multicolumn{3}{c}{All layers} \\
\middlehline
Water path ({g\,m$^{-2}$})  & \multicolumn{3}{c}{[$10^{0.4}, 10^{0.6}, 10^{0.8},
 \ldots , 10^{2.2}$]} \\
Cloud fraction & \multicolumn{3}{c}{[0.05, 0.10, 0.15, \ldots, 1.00]} \\
\bottomhline
\end{tabular}
\end{center}
\end{table}
      For each cloud layer, cloud fraction and water path are varied widely
      whilst the other four parameters are fixed
      (Table~\ref{t-largetune}). In all of the resulting cloud cases, we run
      the radiative transfer code for both standard and elevated \chem{CO_2}
      levels (369\,ppmv and 50\,000\,ppmv), giving 16 million runs in total.

We refer to model runs in which we include clouds in this way as including ``real clouds''. This is meant in the sense that clouds are included in the radiative transfer code in a detailed and physically based manner. This is by contrast to previous models \citep[e.g.][]{kpc-84}, where clouds are represented non-physically by changing surface albedo. 

\subsection{Sensitivity experiment}

      For each cloud case that we have defined (the large ensemble,
      Table~\ref{t-largetune}), we calculate the radiative forcing at the
      tropopause ($\mathbb{F}_{\rm trop}$), to which change in surface
      temperature is proportional. Radiative forcing is the change in net flux (here with increasing $\chem{CO_2}$):
\begin{equation}
\mathbb{F}_{\rm trop} = F_{\rm [trop,high\chem{CO_2}]} - F_{\rm
[trop,std\chem{CO_2}]} \, ,
\end{equation}
where $F$ in each case is the net flux as a sum of longwave and shortwave spectral regions and upward and downward streams of radiation: 
\begin{equation}
F = (F_{\rm lw}^\downarrow - F_{\rm lw}^\uparrow)+(F_{\rm sw}^\downarrow - F_{\rm sw}^\uparrow) \, .
\end{equation}

      We consider two subsets of the large ensemble:
\begin{enumerate}

\item Cloud sets which give energy balance at the TOA. This is the most basic
constraint on a~possible climate. With $|F_{\rm TOA}| {<} 5$\,{W\,m$^{-2}$},
a~subset of 1.0 million cases remains. A~relatively large $|F_{\rm
TOA,std\chem{CO_2}}|$ is allowed as variations in the water path are coarse,
but it is corrected for by calculating radiative forcings so cannot bias the
outcome.

\item Cloud sets which give energy balance at the TOA and are close to
observed longwave and shortwave fluxes at the TOA \citep{tfk-09}. Constraints
are $|F_{\rm TOA,std\chem{CO_2}}| {<} 5$\,{W\,m$^{-2}$}, $95 {<} F\uparrow_{\rm
TOA,SW,std\chem{CO_2}} {<} 115$\,{W\,m$^{-2}$} and $227 {<}
F\uparrow_{\rm TOA,LW,std{CO_2}} {<} 247$\,{W\,m$^{-2}$}). This gives a~subset
of 36\,985 cases.
\end{enumerate}

\begin{figure}
\vspace*{2mm}
\center
\includegraphics[width=8cm]{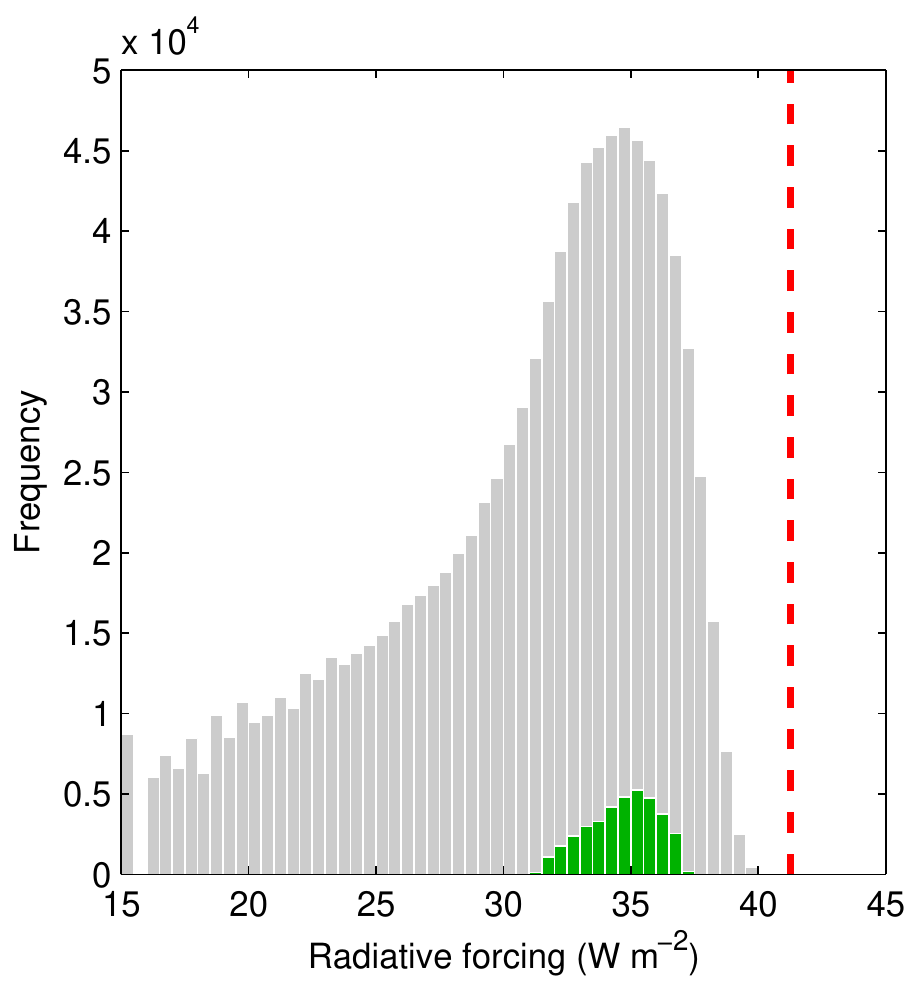}
\caption{Histogram of radiative forcings from two subsets of cloud profiles.
Cloud sets which give energy balance at the TOA (subset~1) in light grey,
cloud sets which give energy balance at the TOA and are close to observed
longwave and shortwave fluxes at the TOA superimposed darker (green online).
Dashed vertical line (red online) shows radiative forcing cloud-free case for
comparison.} \label{f-sens_expt_hist}
\end{figure}
      The distribution of radiative forcings in these two subsets is shown
      relative to the cloud-free radiative forcing of 41.3\,{W\,m$^{-2}$}
      (Fig.~\ref{f-sens_expt_hist}). The maximum radiative forcing from
      subset 1 is 40.2\,{W\,m$^{-2}$}; all physically plausible cloud sets
      give a~smaller radiative forcing than a~cloud-free model. Subset 2 --
      of cloud sets which give Earth-like climate -- has a~mean radiative
      forcing of 34.6 with a~standard deviation of 1.3\,{W\,m$^{-2}$}. The
      radiative forcing from the cloud-free case is 4.9 standard deviations
      above the mean radiative forcing from realistic, Earth-like, clouds.

Note that we perform these runs at the standard solar constant, as for all model runs herein. Given that $\chem{CO_2}$ is not a strong absorber in the shortwave spectral region, selecting a lower solar constant for the sensitivity experiment would not cause any noticeable change to Fig.~\ref{f-sens_expt_hist}. For example, using a solar flux 80\% of the present value yields forcings different by 0.2 to 0.3\,{W\,m$^{-2}$}.

\subsection{Case study selection}

\def\yx{\kern5.5pt}
\def\dt{\kern1.5pt}
\begin{table}[t]
\caption{Cloud properties used in case study. $f_{\rm total} {=} 0.66$.}
\label{t-casestudyprop}
\vskip4mm
\begin{center}
\begin{tabular}{lccc}
\tophline
Property & High & Mid & Low \\
\middlehline
Cloud top (hPa) & 250 & 500 & 700 \\
Cloud base (hPa) & 300 & 600 & 850 \\
Cloud fraction  &  0.25 & 0.25 & 0.40 \\
Water path ({g\,m$^{-2}$}) &  {\yx}20 & {\yx}25 & {\yx}40 \\
Liquid or ice & Ice & Liquid & Liquid \\
Generalised effective size (
\unit{\mu}m)  &  {\yx}75 & -- & -- \\
Effective radius (\unit{\mu}m) & -- & {\yx}11 & {\yx}11  \\
\bottomhline
\end{tabular}
\end{center}
\end{table}
\begin{figure*}
\vspace*{2mm}
\center
\includegraphics{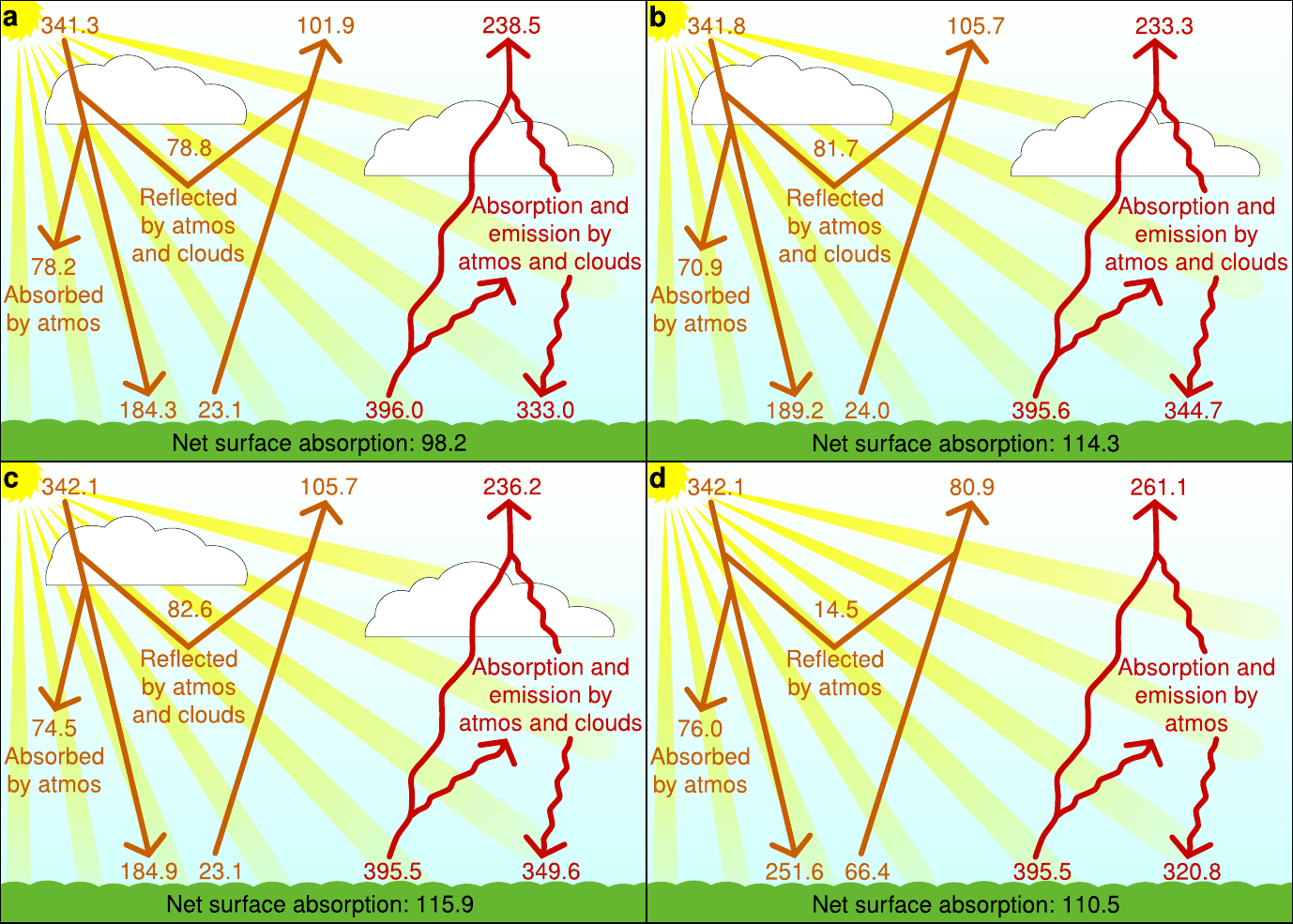}
\caption{Comparison of global annual mean energy budgets: two estimates for
contemporary climate, \textbf{(a)}~\citet{tfk-09}, based on a~composite of
data and \textbf{(b)}~\citet{zrlom-04}, from ISCCP-FD data, compared to
models used in this paper, \textbf{(c)}~case study with real clouds
\textbf{(d)}~cloud-free model.} \label{f-energy_bal_4panel}
\end{figure*}
      As discussed, there are many problems associated with selecting a~set
      of cloud profiles. However, the radiative forcings from \chem{CO_2}
      enhancement in all Earth-like cloud sets are closely grouped
      (Fig.~\ref{f-sens_expt_hist}) and the mean of these is significantly
      different from the cloud-free case. This justifies definition of
      a~case study which can be used to represent Earth's clouds. To do this
      from subset 2, we additionally constrain cloud fractions (each layer and the
      resultant total) and water paths of each layer to be close to
      climatological values, optimising for agreement with longwave and
      shortwave fluxes at the TOA. We found that, whilst shortwave fluxes
      could be found that were in good agreement with climatological values,
      the outgoing longwave fluxes were slightly too high in all cases from
      ensemble 2. Increasing the height of the clouds by 50\,hPa gives
      a~better fit for longwave fluxes. Case study cloud properties are
      given in Table~\ref{t-casestudyprop} and the radiative outcome in
      Fig.~\ref{f-energy_bal_4panel}c.

\subsection{Cloud-free case}

      In order to compare calculated cloudy and cloud-free radiative forcings we need a~cloud-free model as a comparison case. To generate
      this, we follow \citet{kpc-84} and tune the surface albedo of the GAM
      profile to achieve energy balance at the top of the atmosphere for
      a~clear sky profile. The required surface albedo is 0.264.

\section{Real clouds and cloud-free model compared}
\label{s-kastingeval}

      First, consider the energy budget at standard conditions relative to
      observational climatology (Fig.~\ref{f-energy_bal_4panel}). Our case
      study with real clouds (Fig.~\ref{f-energy_bal_4panel}c) is in very
      good agreement with observational climatology
      (Fig.~\ref{f-energy_bal_4panel}a,b). By contrast, almost all of the
      variable fluxes in the cloud-free model
      (Fig.~\ref{f-energy_bal_4panel}d) are markedly different; omitting
      clouds means that the global energy budget is not properly
      represented. Overall in the cloud-free model, more absorption of solar
      radiation (only 81\,{W\,m$^{-2}$} of outgoing shortwave radiation is reflected
      rather than 106\,{W\,m$^{-2}$}, a~lower overall planetary albedo) is
      balanced by a~weaker greenhouse effect (with an elevated outgoing
      longwave flux of 261\,{W\,m$^{-2}$} rather than 236\,{W\,m$^{-2}$} and
      depressed downward longwave at the surface).

      Whilst no 1-D model can perfectly represent global climate, our real cloud case study, which is constrained by observational cloud climatology, gives
      good agreement with the observed energy budget. This justifies using it as an internal standard, against which
      the cloud-free model can be compared. 

\begin{figure}
\vspace*{2mm}
\center
\includegraphics[width=8cm]{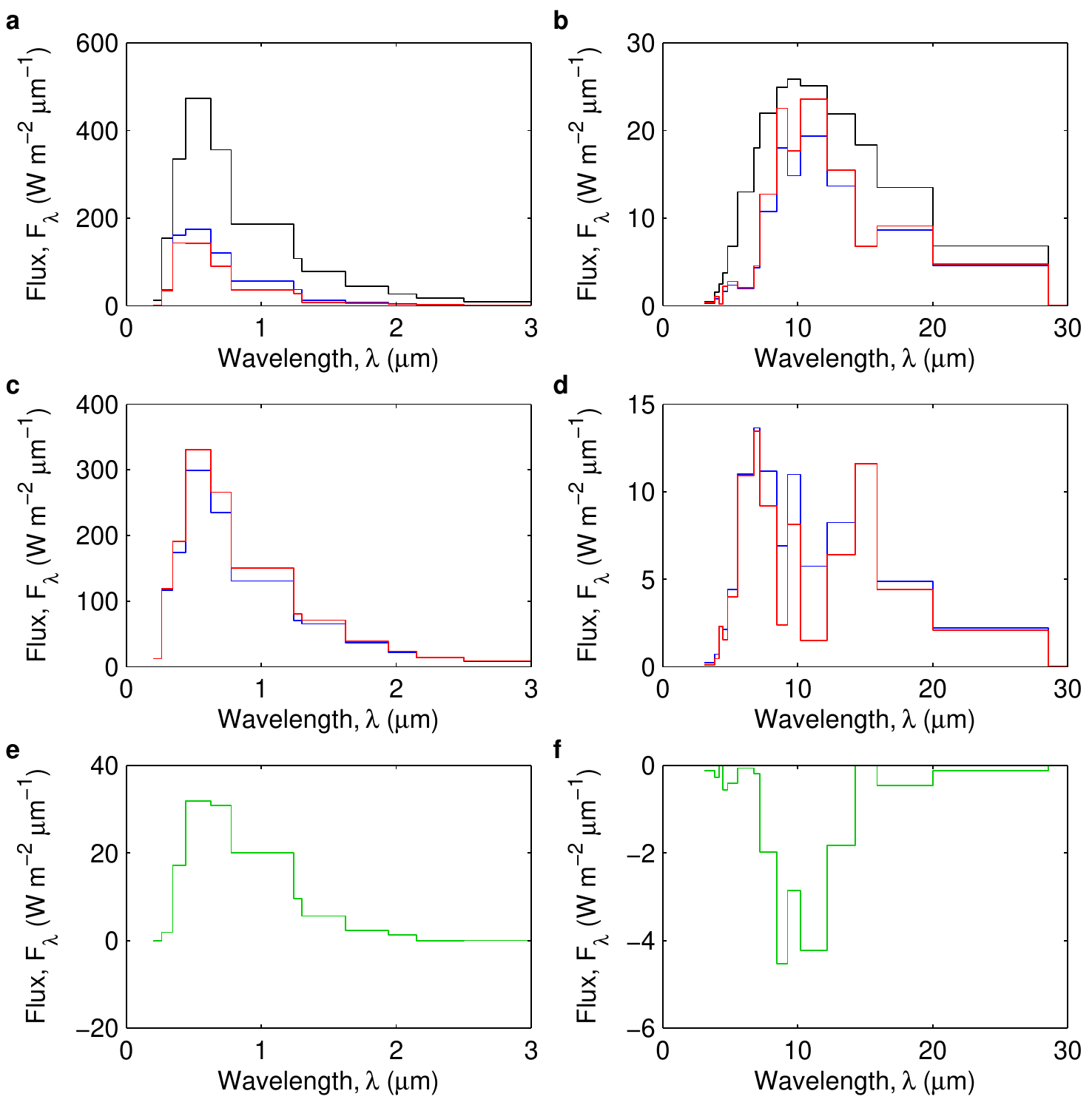}
\caption{Comparison of spectrally resolved energy budgets in Real Cloud
(RC, blue) and Cloud Free (CF, red) models. Black lines are for both cases.
Green is for differences (CF${-}$RC). \textbf{(a)}~$F\downarrow_{\rm SW}^{\rm
TOA}$ for both cases in black, $F\uparrow_{\rm SW}^{\rm TOA}$ in colours.
\textbf{(b)}~$F\uparrow_{\rm LW}^{\rm surf}$ for both cases in black,
$F\uparrow_{\rm LW}^{\mathrm{TOA}}$ in colours. \textbf{(c)}~Absorption of
solar radiation: $A {=} F\downarrow_{\rm SW}^{\rm TOA} {-} F\uparrow_{\rm
SW}^{\rm TOA}$ \textbf{(d)}~Greenhouse effect: $G {=} F\uparrow_{\rm LW}^{\rm
surf} {-} F\uparrow_{\rm LW}^{\rm TOA}$ \textbf{(e)}~Difference in solar
absorption: $D_A {=} A({\rm CF}) {-} A({\rm RC}$) \textbf{(f)}~Difference in
greenhouse effect: $D_G {=} G({\rm CF}) {-} G({\rm RC}$).}
\label{f-spec_out_std_pub}
\end{figure}
Again at standard conditions,
      compare the spectrally resolved fluxes
      (Fig.~\ref{f-spec_out_std_pub}). In the shortwave, the difference in
      adsorption between cloud-free and real cloud models
      (Fig.~\ref{f-spec_out_std_pub}e) has the same shape as the Planck
      function of solar radiation (Fig.~\ref{f-spec_out_std_pub}c). This is
      because the surface albedo is constant with wavelength by definition
      and the wavelength dependence of cloud scattering is weak. Rayleigh
      scattering is spectrally dependent (short wavelengths are
      preferentially scattered), but this is a~small term
      (14.5\,{W\,m$^{-2}$} in the cloud-free case). By contrast, in the
      longwave, there is strong spectral dependence in the differences
      between the real cloud and cloud-free models. The cloud-free model has
      a~weaker greenhouse effect than real clouds in the water vapour window
      region. Whilst other spectral regions are optically thick (with
      gaseous absorption by water vapour and carbon dioxide dominating) the
      water vapour window is optically thin and the cloud greenhouse is
      important.

\begin{figure}
\vspace*{2mm}
\center
\includegraphics[width=8cm]{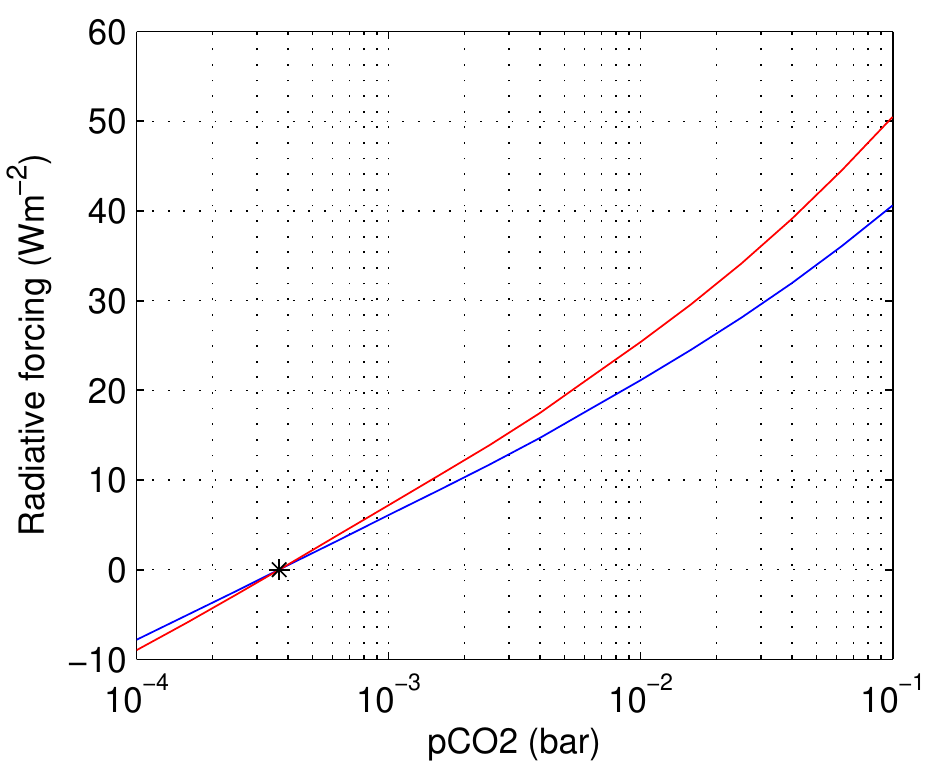}
\caption{Radiative forcing with increasing $p\chem{CO_2}$. Real clouds in
blue and cloud-free in red. Present $p\chem{CO_2}$ marked (\ding{83}).}
\label{f-casestudy_rftrop_co2_pub}
\end{figure}
      Now consider the effect of changing \chem{CO_2} concentration
      (Fig.~\ref{f-casestudy_rftrop_co2_pub}). Radiative forcing is strongly
      overestimated by the cloud-free model relative to our real cloud case study; to
      produce a~given radiative forcing, twice as much \chem{CO_2} is needed
      with the real cloud case study than is indicated by the cloud-free model.

\begin{figure}
\vspace*{2mm}
\center
\includegraphics[width=8cm]{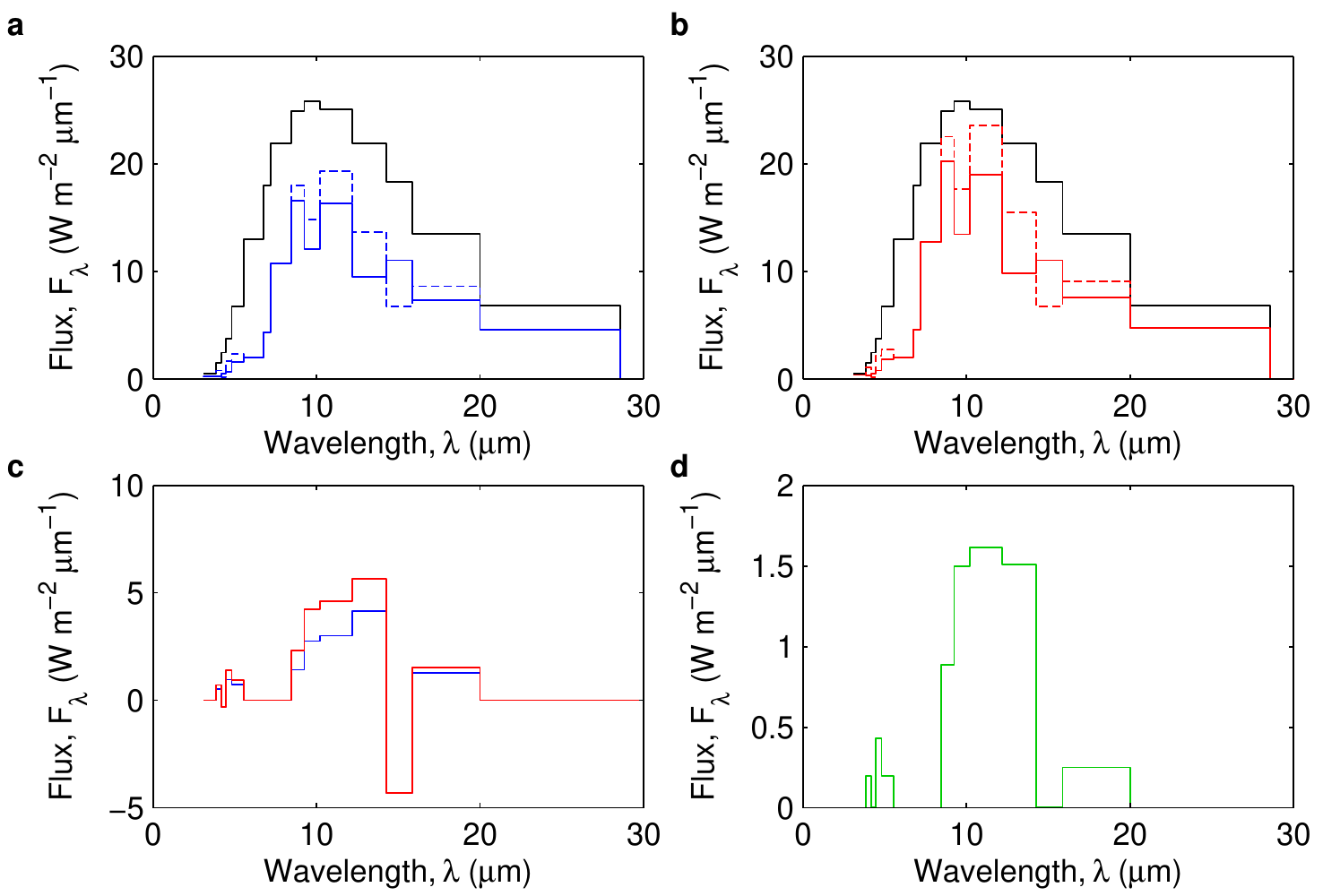}
\caption{Comparison of spectrally resolved longwave forcings for increase
from standard to 50\,000\,ppmv \chem{CO_2} in real cloud (RC, blue) and
cloud-free (CF, red) models. Green is for differences in cases (CF${-}$RC)
and black is for fluxes common between cases. \textbf{(a)}~RC:
$F\uparrow_{\rm LW}^{\rm surf}$ in black, $F\uparrow_{\rm LW}^{\rm TOA}$
dashed blue for standard \chem{CO_2} and solid blue for elevated \chem{CO_2};
\textbf{(b)}~CF: $F\uparrow_{\rm LW}^{\rm surf}$ in black, $F\uparrow_{\rm
LW}^{\rm TOA}$ dashed red for standard \chem{CO_2} and solid red for elevated
\chem{CO_2}; \textbf{(c)}~greenhouse forcing from increased \chem{CO_2}:
$\mathbb{G} {=} G({\rm High} \chem{CO_2}) {-} G({\rm Std} \chem{CO_2}$) where $G {=} F\uparrow_{\rm LW}^{\rm
surf} {-} F\uparrow_{\rm LW}^{\rm TOA}$;
\textbf{(d)}~difference in greenhouse forcing: $D_\mathbb{G} {=}
\mathbb{G}({\rm CF}) {-} \mathbb{G}({\rm RC}$).} \label{f-spec_out_diffco2_pub_alt}
\end{figure}
\begin{figure*}
\vspace*{2mm}
\center
\includegraphics[width=15cm]{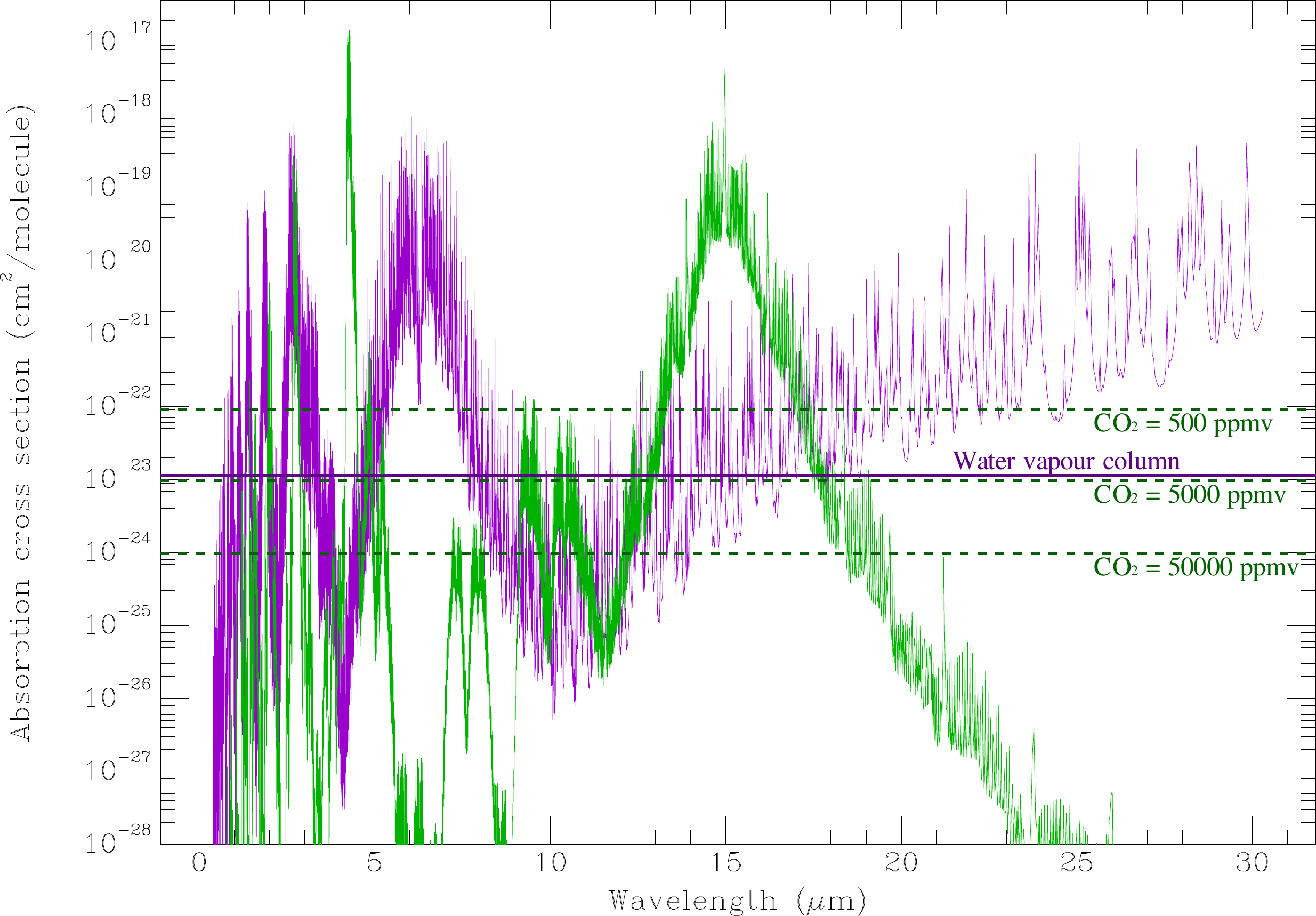}
\caption{Absorption cross sections for \chem{CO_2} (green) and water vapour
(purple) from HITRAN (shown at 900\,hPa and 285\,K). Horizontal lines
indicate the cross section for which the gas has an optical depth of unity,
solid purple for the GAM water vapour column, dashed green for various
\chem{CO_2} concentrations. The column depth of the atmosphere is
$2.1{\times}10^{25}$\,molecules\,cm$^{-2}$. } \label{f-plt_um_annotated}
\end{figure*}
      The radiative forcing in the longwave is an order of magnitude larger
      than the radiative forcing in the shortwave region, so we focus on the
      longwave region when comparing spectrally resolved forcing
      (Fig.~\ref{f-spec_out_diffco2_pub_alt}). The greenhouse effect with
      real clouds is stronger at standard conditions than the cloud-free
      model (inclusion of cloud greenhouse). However, the greenhouse forcing (increase in strength of the greenhouse effect), is larger in the cloud-free model. This is true across all bands where
      \chem{CO_2} imparts a~greenhouse effect and is most important in the
      water vapour window. Here, the atmosphere is optically thin in the
      absence of clouds, so the effect of increasing \chem{CO_2} is large
      even though its absorption lines are weak
      (Fig.~\ref{f-plt_um_annotated}). With clouds, these regions will be
      optically thicker initially so increasing \chem{CO_2} has less of an
      effect.

      At 15\,{\unit{\mu}m}, increasing \chem{CO_2} causes increased longwave
      emission. This is due to increased emission in the stratosphere and is
      therefore unaffected by tropospheric clouds.

      Our GAM profile includes \chem{O_3} which absorbs at
      9.5\,{\unit{\mu}m} and 9.7\,{\unit{\mu}m}. This would be absent in the
      anoxic Archean atmosphere, making the water vapour window optically
      thinner. The overestimation of forcing by the cloud-free model is,
      therefore, likely larger than suggested here and even more \chem{CO_2}
      would actually be needed to cause equivalent warming.

The other perturbation to consider is the change in incoming solar flux. The cloud-free model absorbs a higher proportion of the incoming solar flux (has a lower planetary albedo), so will have a proportionately larger response to changing solar flux.  For an 20\% decrease in solar flux, representative of the late Archean, the decrease in absorbed solar flux is 52.3\,{W\,m$^{-2}$} for the cloud-free model compared to 47.2\,{W\,m$^{-2}$} for the real-cloud case study. Taking the Archean to be lower solar flux but higher \chem{CO_2}, this error is of opposite sign to the error in radiative forcing from increased \chem{CO_2} and around half the magnitude. Whilst these errors could be said to partially offset in these conditions, reliance on errors of opposing sign is not strong. 

\section{Variation of cloud and surface properties}
\label{s-cloudpropvar}

      The problem of cloud feedback on climate change is notoriously
      difficult. We do not attempt to address this in full; rather, we
      explore how variations in cloud amounts and properties could affect
      climate. In all cases here, our baseline case is the real cloud case study and
      we consider the radiative effect of changes in cloud or surface
      properties. As comparison values, if we increase or decrease the
      humidity in the model profile by 10\% (50\%), the radiative forcings
      are 1.7\,{W\,m$^{-2}$} (7.8\,{W\,m$^{-2}$}) and
      ${-}$1.9\,{W\,m$^{-2}$} (${-}$11.2\,{W\,m$^{-2}$}),
      respectively. A~fainter sun in the late Archean is equivalent to
      a~forcing of around ${-}$50\,{W\,m$^{-2}$} (assuming a~planetary
      albedo of 0.3).

\subsection{Surface albedo}

      In the cloud-free model, the use of a~non-physical surface albedo
      means that real changes in surface albedo cannot be considered. This
      limitation is removed with explicit clouds. We limit discussion here
      to changes in surface albedo not from ice, though the use of
      a~physically realistic surface with RC means that a~parameterised
      ice-albedo feedback could be included in 1-D climate
      models, a~significant improvement on the status quo.

\begin{figure}
\vspace*{2mm}
\center
\includegraphics{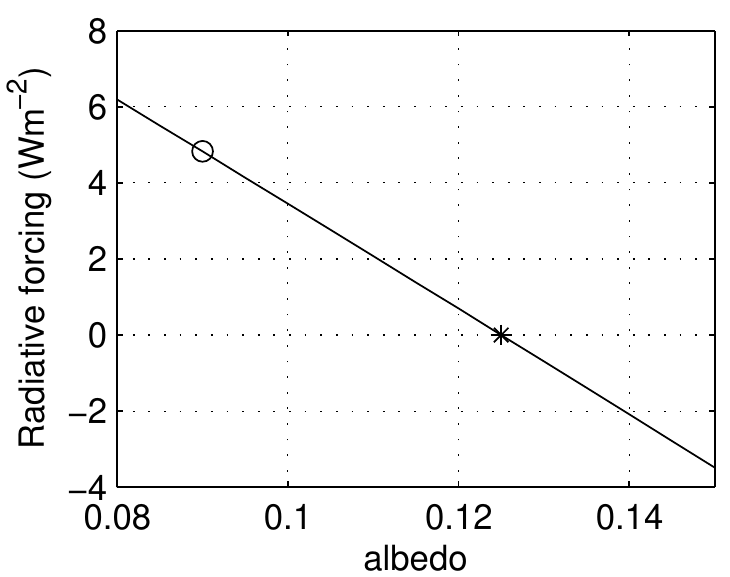}
\caption{Radiative forcing with changed surface albedo. (\ding{83}) is the
case study and (~$\circ$~) is the end-member case of an ocean covered planet.
\label{f-cloudprop_surfalb_pub}}
\end{figure}
      The surface albedo we use of 0.125 represents a~weighted average of
      land (0.214) and ocean (0.090) albedos \citep{tfk-09}. Continental
      volume is generally thought to have increased over time, with perhaps
      up to 5\% of the present amount at the beginning of the Archean and
      20\%--60\% of the present amount by the end of the Archean
      \citep[e.g.][]{hk-06}. In Fig.~\ref{f-cloudprop_surfalb_pub} we
      consider a~range of variation of surface albedo appropriate for
      a~changed land fraction. For the end-member case relevant to the
      Archean of a~water-world, the radiative forcing is
      4.8\,{W\,m$^{-2}$}. Without land, relative humidity would likely be
      higher, contributing extra forcing.\vspace*{1.5mm}

\subsection{Cloud fraction and water path}\vspace*{1mm}

      There are more clouds over ocean than land
      (Fig.~\ref{f-cloudfrac_4pcol}). The zonally uninterrupted Southern
      Ocean is especially cloudy. One might therefore expect that when there
      was less land there would have been more cloud, and more still if
      there was a~greater extent of zonally uninterrupted ocean. Comparison
      of the Northern and Southern Hemispheres of Earth
      (Fig.~\ref{f-cloudfrac}), the former having a~higher land fraction,
      may be indicative of the minimum expected degree of variation. The
      Southern Hemisphere has 20--50\% greater cloud fraction in each layer
      than the Northern Hemisphere.

\begin{figure*}
\vspace*{2mm}
\center
\includegraphics{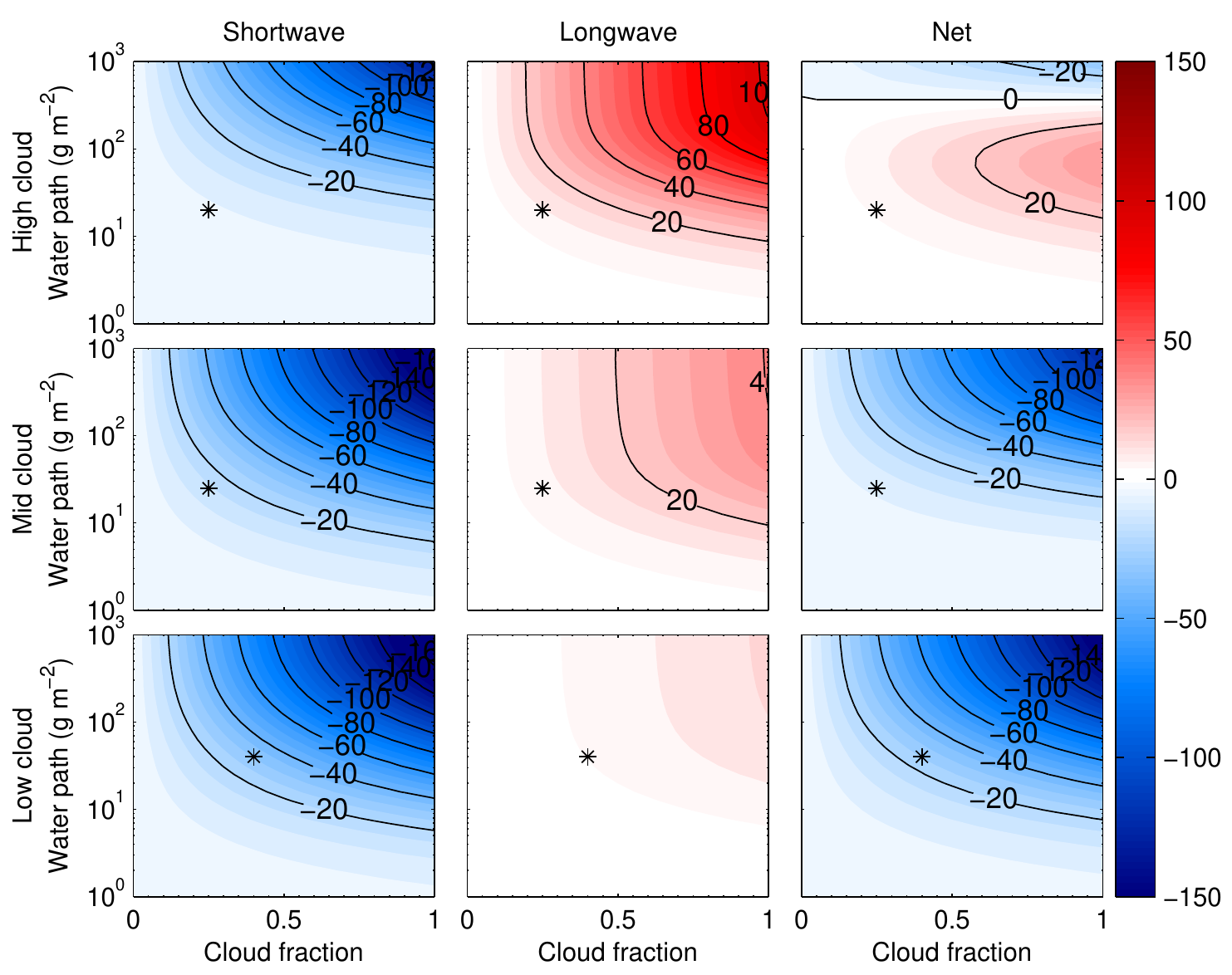}
\caption{Cloud radiative forcing with cloud fraction and water path relative
to no cloud in that layer. For each cloud layer, these properties are varied
whilst the clouds in other layers remain fixed at the case study values,
marked (\ding{83}). Particle sizes, ice/water ratio and height are as case
study. Colour/contour scale is in {W\,m$^{-2}$}. For comparison, resolution
of the late Archean FYSP would require a~forcing of approximately
50\,{W\,m$^{-2}$}. \label{f-cloudprop_arrayexpt}}
\end{figure*}
      We consider a~wide range of water paths, from optically thin to thick
      clouds, and fractional cloud cover from zero to 1 for each cloud
      layer. In Fig.~\ref{f-cloudprop_arrayexpt} we show the radiative
      forcing from these clouds relative to no cloud in the given layer. The
      competing shortwave and longwave effects of changing clouds can
      readily be seen. Increasing fraction or water path causes a~negative
      forcing in the shortwave region (more reflection) but a~positive
      forcing in the longwave (greenhouse effect). The greenhouse effect
      operates by absorption of thermal radiation emitted by a~warm surface
      followed by emission at a~lower temperature. Therefore the magnitude
      of changes in the greenhouse effect varies with cloud height, as
      higher clouds are colder. For low and mid level clouds, shortwave
      effects dominate and increasing cloud fraction or thickness will cause
      a~net negative forcing (cooling the planet). For high clouds,
      shortwave and longwave effects are of similar magnitudes so the
      character of the net response is more complicated. For water paths
      less than 350\,{g\,m$^{-2}$}, high clouds cause a~net positive forcing
      (greenhouse warming the planet). The converse is true above
      350\,{g\,m$^{-2}$}, but such high water paths would typically
      correspond to deep convective clouds, not high clouds (cirrus or
      cirrostratus) \citep{rs-99}. Positive forcing is maximum for
      $\sim$70\,{g\,m$^{-2}$} high clouds.

\subsection{Cloud particle size}
\label{s-partsize}

      Cloud particle size depends very strongly on the availability of cloud
      condensation nuclei (CCN). Whilst the global mean droplet size is
      11\,{\unit{\mu}m}, this is biased by smaller droplets over land
      (average 8.5\,{\unit{\mu}m}), where there are more CCN than over the
      ocean (average 12.5\,{\unit{\mu}m}). Over the ocean, around half of
      CCNs are presently derived from oxidation products of biogenic
      dimethyl sulphide (DMS), especially sulphuric acid (there are various oxidation pathways of DMS \citep[e.g.][]{vonglasow-crutzen-04}, but only sulphuric acid can cause nucleation of new droplets \citep{krei-sein-88}). The climatic
      feedbacks involving DMS \citep{claw-87} have been subject of long
      debate. Whilst DMS is prevalent today due to production by eukaryotes,
      other biogenic sulphur gases are produced by bacteria, in particular
      hydrogen sulphide (\chem{H_2S}) and methyl mercaptan (\chem{CH_3SH})
      \citep{kettle-ea-01}. These will react chemically to form sulphates,
      which will provide CCN.

      We do not delve deeply into CCN feedbacks here, but accept that
      various changes in the Earth system (e.g.\ atmospheric oxidation
      state, sulphur cycle, volcanic fluxes, biological fluxes) may well
      have changed CCN availability. Fewer CCN give larger cloud drops,
      which should both rain out quicker (so less cloud) and be less
      reflective. Conversely, more CCN give more extensive and more
      reflective clouds.

\begin{figure*}
\vspace*{2mm}
\center
\includegraphics{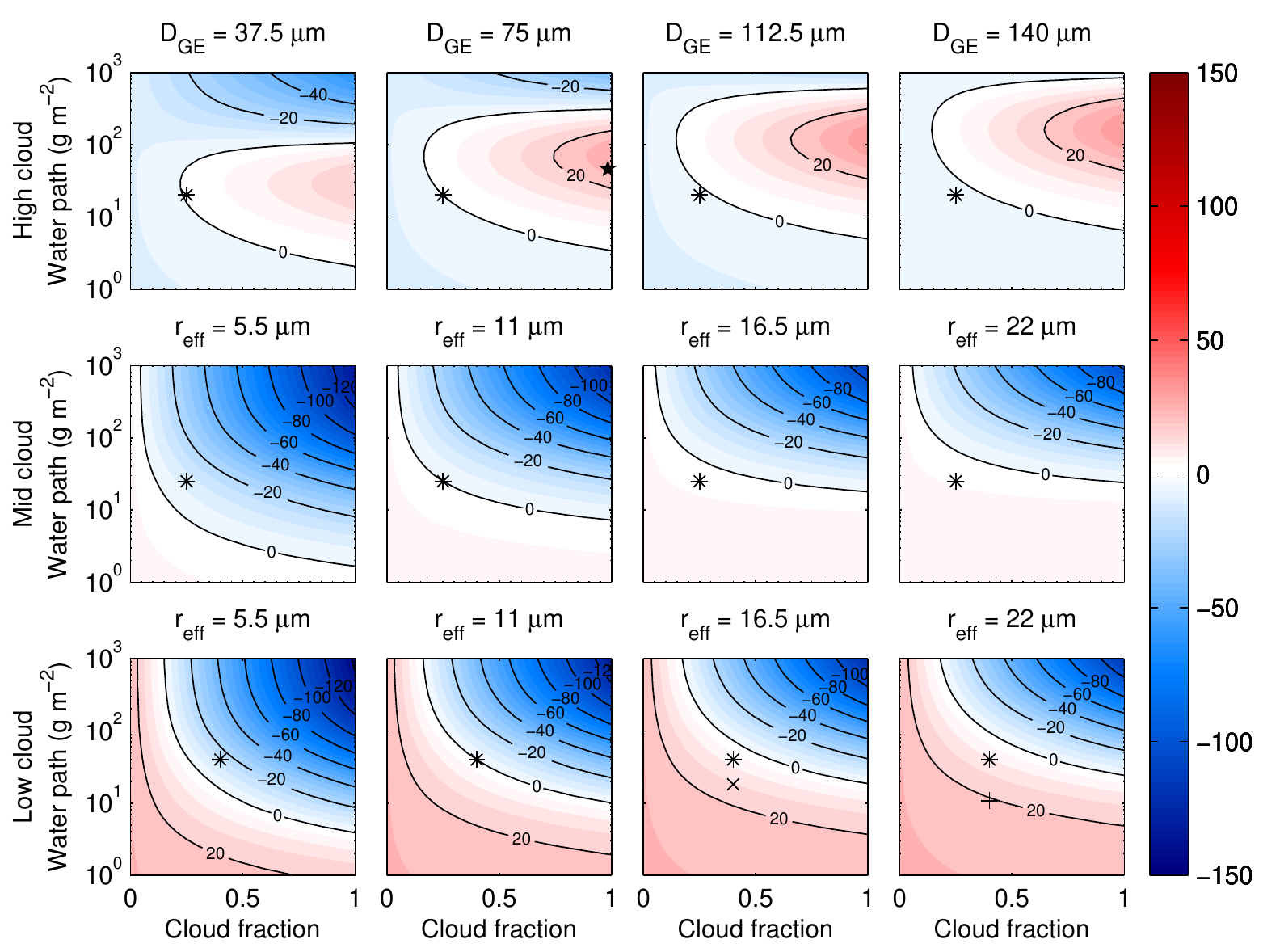}
\caption{Change in net cloud radiative forcing with cloud particle size,
across a~range of cloud fractions and water paths. Particle size is varied
for each layer independently (values in subplot titles), whilst all
properties for other cloud layers remain as case study. The change is shown
relative to the case study (so panels for $r_{\rm eff} {=} 11$\,{\unit{\mu}m}
and $D_{\rm GE} {=} 75$\,{\unit{\mu}m} contain the same information as
Fig.~\ref{f-cloudprop_arrayexpt} net fluxes). Cloud fraction and water path
for case study are marked (\ding{83}); values at the point of these markers
are for changing particle size only, values elsewhere in each panel are for
changing water path or fraction too. Markers (${\times}$) and (${+}$) refer
to reduction in water path by factors of 2.2 and 3.7, respectively, for
comparison to \citet{rosing-ea-10}, as discussed in the text. Marker
(\ding{83}) corresponds to the relatively thick and maximum extent clouds
invoked by \citet{ron-lindzen-10}. Colour/contour scale is in {W\,m$^{-2}$}.}
\label{f-cloudprop_reffdgearrayexpt_alt1}
\end{figure*}
      We consider the effect of changing liquid droplet size by factors of
      0.5, 1.5 and 2 relative to the case study ($r_{\rm eff} {=}
      11$\,{\unit{\mu}m}) and ice particle size by factors of 0.5, 1.5 and
      1.87 relative to the case study ($D_{\rm GE} {=} 75$\,{\unit{\mu}m}; the
      maximum of the parameterisation used is 140\,{\unit{\mu}m}). In
      Fig.~\ref{f-cloudprop_reffdgearrayexpt_alt1}, we show the net
      (shortwave plus longwave) radiative forcing from changing particle
      size for all water paths and fractions. The effect is strongest for
      low clouds. With no change to cloud fraction or water path, increasing
      $r_{\rm eff}$ by 50\% gives a~forcing of 7.5\,{W\,m$^{-2}$} and
      doubling $r_{\rm eff}$ gives a~forcing of
      10.4\,{W\,m$^{-2}$}. Decreasing $r_{\rm eff}$ by 50\% gives
      a~forcing of ${-}$13.6\,{W\,m$^{-2}$}.

      Satellite observations of the modern ocean \citep{breon-ea-02}
      suggests a~limit on how large droplet size actually becomes in
      nature. Particle size is rarely larger than 15\,{\unit{\mu}m}, even in
      the remotest and least productive regions of the ocean. Here, the DMS
      flux is low and remaining CCN derive from abiological sources (e.g.\
      sea spray). $r_{\rm eff} {=} 15$\,{\unit{\mu}m} can then be seen as
      the baseline case for lower CCN availability, corresponding to a~36\%
      size increase relative to present day mean (20\% relative to present
      day ocean).

      If there was a~larger CCN flux, the droplet size for clouds over land
      ($r_{\rm eff} {=} 8.5$\,{\unit{\mu}m}, 23\% less than mean) is an
      indicator of likely droplet size.

      Larger droplets will rain out more effectively, but model
      representations of this feedback vary dramatically
      \citep{penner-ea-06,kp-08}. For the case of $r_{\rm eff} {=}
      17$\,{\unit{\mu}m} droplets over the ocean, \citet{kp-08} choose
      a~mid-strength assumption of this feedback, implying a~decrease of
      water path by a~factor of 2.2. This is marked (${\times}$) in the low
      cloud, 16.5\,{\unit{\mu}m} panel of
      Fig.~\ref{f-cloudprop_reffdgearrayexpt_alt1}; the radiative forcing is
      then 15.4\,{W\,m$^{-2}$}, twice that of solely increasing droplet
      size. Clearly, an increased precipitation feedback is of first order
      importance and must be treated carefully in any model addressing the
      climatic effect of changed particle size.

\subsection{Cloud height}

\begin{figure}
\vspace*{2mm}
\center
\includegraphics[width=8cm]{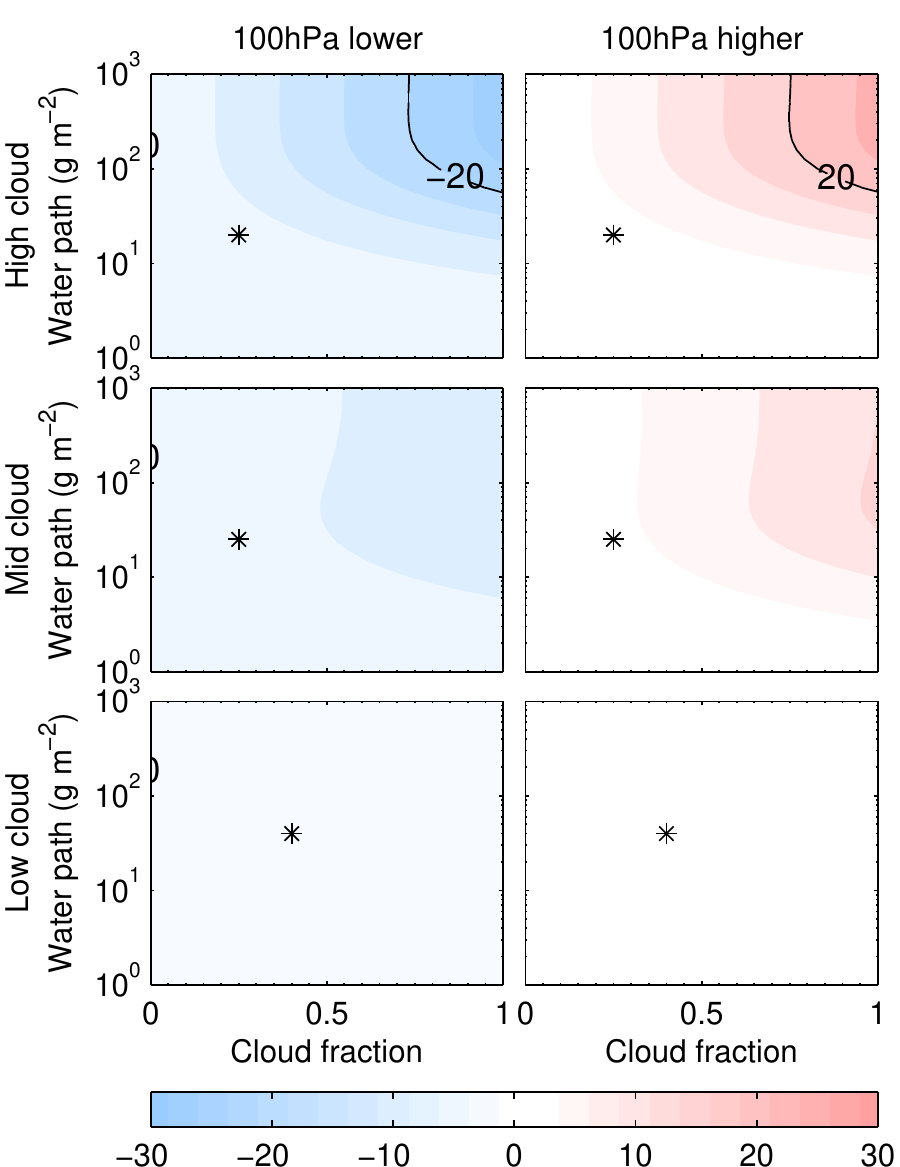}
\caption{Cloud radiative forcing with changed cloud height relative to
standard height clouds (see Fig.~\ref{f-cloudprop_arrayexpt}). Colour/contour
scale is in {W\,m$^{-2}$}.} \label{f-cloudprop_arrayheight_alt}
\end{figure}
      To test the sensitivity to cloud height, each cloud layer is raised or
      lowered 100\,hPa and the forcing is calculated relative to the
      standard heights for all water paths and fractions. As temperature
      decreases with height, higher clouds emit at a~lower temperature. It
      is this longwave effect which is dominant. There are only small
      changes in the shortwave effect, due to changed path length above the
      cloud (a~greater path length means decreased insolation due to more
      Rayleigh scattering in the overlying atmosphere). In
      Fig.~\ref{f-cloudprop_arrayheight_alt} we show only the net
      forcing. For the high clouds in the case study, which cover
      one-quarter of the sky and have a~water path of 20\,{g\,m$^{-2}$}, the
      effect of raising them is relatively small (2.8\,{W\,m$^{-2}$}). There
      is a~larger forcing from raising clouds which are thicker or cover
      more of the sky initially; the greater the radiative longwave effect
      of the cloud at its standard height, the greater the effect of
      changing it height would be.

      Changes in the temperature--pressure structure of the atmosphere might
      have induced changes in clouds. The Archean atmosphere was
      anoxic and did not have an ozone layer
      \citep{kd-80,glw-06}. Consequently, there would likely not have been
      a~strong stratospheric temperature inversion, and deep atmospheric
      convection may have reached higher altitudes, where the atmosphere is
      colder. A~major source of high clouds is detrainment of cirrus from
      deep convective clouds. Where detrainment is due to wind sheer, this
      could then result in higher clouds. Conversely, without an inversion
      a~the tropopause, cumulonimbus incus (anvil shaped clouds) will not
      form. As the forcing from raising the high clouds in the case study is
      small, other climatic effects might be larger (loss of ozone as
      a~greenhouse effect and lower stratospheric emission
      temperature). Also, the pressure of Archean atmosphere was likely not
      1\,bar. Not only was there no oxygen (0.21\,bar today), but the
      nitrogen inventory was likely different \citep{g-n2-09}. Varying
      pressure would have changed both the lapse rate and tropopause
      pressure \citep{g-n2-09}.

\section{Evaluating cloud-based proposals to resolve the Faint Young Sun Paradox}
\label{s-fys_resol_eval}\vspace*{1.5mm}

\subsection{Increased cirrus}\vspace*{1.5mm}

      \citet{ron-lindzen-10} proposed that near total coverage of cirrus
      clouds could resolve the FYSP. Their proposed mechanism is that the
      planet would be colder and have lower sea surface temperatures, which
      would give more cirrus coverage \citep[the controversial \qut{iris}
      hypothesis of][]{lindzen-ea-01}, acting as a~strong negative feedback
      on temperature. The first premise here, of colder temperatures, is
      contrary to the geological record; this suggests less frequent
      glaciation through the Archean and Proterozoic than in the
      Phanerozoic, not more. The second premise, of strong cloud feedback,
      is based on a~statistical relationship for Earth's tropics
      \citep{lindzen-ea-01} the authenticity of which has been questioned
      \citep[e.g.][]{hartmann-michelsen-02,chambers-ea-02}. Application to
      very cold temperatures requires an extreme and unverifiable
      extrapolation. \citet{ron-lindzen-10} describe the high level clouds
      they use as \qut{thin cirrus}; we note that the clouds they use
      actually have twice the water path of our standard high clouds. In
      their sensitivity tests, using a~thinner high level clouds gives
      a~weaker effect.

      Here, we consider what would be required of cirrus or other high level
      clouds for them to resolve the FYSP. Informed by the experiments
      above, we construct an optimum cirrus cloud for warming: relative to
      our case study we make it 3.5 times thicker (a~water path of
      70\,{g\,m$^{-2}$}) and make it cover the whole sky, not just
      one-quarter (similar to the suggestion of
      \citealp{ron-lindzen-10}). This gives a~forcing of 29.0\,{W\,m$^{-2}$},
      insufficient to counter the ${\sim}$50\,{W\,m$^{-2}$} deficit from the
      FYSP. If, in addition, we raise the cloud by 100\,hPa (base at 200\,hPa,
      making the cloud 14\,K colder) the total radiative forcing becomes
      50.7\,{W\,m$^{-2}$}.

      In principle, high clouds can resolve the FYSP. In practice, the
      requirement for total high level cloud cover seems implausible and the
      requirement that the clouds are higher (colder) is difficult to
      justify.  That it takes an extreme end-member case to provide only
      just enough forcing to resolve the FYSP suggests that resolution with
      enhanced cirrus only is not a~strong hypothesis.

\subsection{Decreased stratus}

      \citet{rosing-ea-10} propose that there were less CCN available in the
      Archean, due to lower DMS emissions prior to the oxygenation of the
      atmosphere and widespread occurrence of eukarya. They suggest an
      increase in droplet size from 12\,{\unit{\mu}m} to 20 or
      30\,{\unit{\mu}m}. Even over unproductive regions of today's oceans,
      the effective radius of cloud particles rarely exceeds
      15\,{\unit{\mu}m} \citep{breon-ea-02}, so it is difficult to see how
      such large effective radii could be justified. Larger droplets lead to
      more rain, so should make clouds thinner. To account for this,
      \citet{rosing-ea-10} arbitrarily decrease the liquid water path of
      their stratus clouds by a~factor of 3.7, which is at the high end of
      likely decreases \citep{penner-ea-06}. Even with these very strong
      assumptions, their model temperature is continually below the present
      temperature before 2\,Ga.

      In our framework of radiative forcings, the effects of changing
      effective radius and cloud water path are shown in
      Fig.~\ref{f-cloudprop_reffdgearrayexpt_alt1}. For the strong but
      arguably plausible case (discussed in Sect.~\ref{s-partsize}) of
      doubling the effective radius and decreasing water path by a~factor of
      2.2 gives a~radiative forcing of 15.4\,{W\,m$^{-2}$}. For the yet
      stronger case of doubling the effective radius from 11\,{\unit{\mu}m}
      to 22\,{\unit{\mu}m} and decreasing cloud water path by a~factor of
      3.7, the radiative forcing is 20.5\,{W\,m$^{-2}$}. Removing low cloud
      entirely gives a~forcing of 25.3\,{W\,m$^{-2}$}. We therefore conclude
      that reducing stratus cannot by itself resolve the FYSP.

A separate hypothesis \citep{shaviv-03,svensmark-07} proposes less stratus on early Earth due to fewer galactic cosmic rays being incident on the lower troposphere. The underlying hypothesis is of a correlation between galactic cosmic ray incidence and stratus amount, through CCN creation due to tropospheric ionization \citep{svensmark-friis-97,svensmark-07}. This hypothesis has been refuted \citep[e.g.][and references therein]{sun-bradley-02,lockwood-frohlich-07,kris-ea-08,bailerjones-09,calogovic-ea-10,kulmala-ea-10}: galactic cosmic rays cause the formation of at most 10\% of CCNs and there is no correlation between galactic cosmic ray incidence and cloudiness. Also of note is that \citet{shaviv-03} requires a highly non-standard climate sensitivity to force his model. In the sensitivity test where he uses a more standard climate sensitivity, it results in a mean temperature $\sim 0^\circ$C during the Archean. Even if the underlying hypothesis had not been refuted, the same arguments as above would apply: plausible decreases in stratus are insufficient to resolve the FYSP.

\conclusions

      When calculating radiative forcing from increased greenhouse gas
      concentrations, we find that omitting clouds leads to a~systematic
      overestimate relative to models in which clouds are included in a physically based manner. With 0.1\,bar \chem{CO_2} (the
      relevant quantity for a \chem{CO_2} based resolution to the Faint
      Young Sun Paradox in the late Archean) the overestimate of radiative
      forcing from modelling without clouds approaches 10\,{W\,m$^{-2}$} in
      our model, equivalent to the clear-sky forcing from 100\,ppm
      \chem{CH_4}. As the radiative transfer code we use underestimates
      forcing from \chem{CO_2} at this level, and we include \chem{O_3} in
      our profile, the difference between the real cloud case study and the
      cloud-free model here must be seen as a~lower bound on the error from
      omitting clouds. For other greenhouse gases, especially those which
      absorb strongly in the water vapour window, the overestimation by
      a~cloud-free model would likely be larger. This would affect
      calculations of the warming by methane and ammonia and of recently
      proposed Archean greenhouse gases, ethane \citep{hdkk-08} and OCS
      \citep{ueno-ea-09}.

      The question of what direct effect clouds might have is a~more
      interesting and difficult one. We can address this best by considering
      what radiative forcing can be generated in both the shortwave and
      longwave spectral regions by changing cloud physical properties, and
      whether such changes in cloud physical properties can be justified.

      For solar radiation (shortwave), low level stratus clouds have the
      greatest effect. Removing them from the model entirely gives a~forcing
      of 25\,{W\,m$^{-2}$}. Even this end-member falls short of the
      50\,{W\,m$^{-2}$} that is needed to resolve the FYSP. A~more plausible
      combination of reduced fraction and water path and increased droplet
      size would give a~maximum forcing of 10--15\,{W\,m$^{-2}$}. However,
      suitable justification for these changes does not come
      easily. \citet{rosing-ea-10} asserted that DMS fluxes would be low in
      the Archean, but there may well have been other biological and
      chemical sources for the sulphuric acid on which water condenses (DMS
      is a~precursor to this). For example, methyl mercaptan is produced
      abundantly by bacteria \citep{kettle-ea-01}. Observations of clouds
      show that the effective radius rarely becomes larger that
      15\,{\unit{\mu}m} \citep{breon-ea-02}, which implies that regionally
      low CCN flux does not lead to very large droplets. If there were less
      land early in Earth's history and more zonally uninterrupted ocean,
      one might expect there to be more cloud rather than less (similar to
      how there is greater cloud fraction in the Southern Hemisphere than
      the Northern Hemisphere). 
Also relating to low cloud, \citet{shaviv-03} and \citet{svensmark-07} contend that less galactic cosmic rays were incident on the troposphere during the Archean and this would have led to less stratus. However, the underlying hypothesis for this has been refuted.

      For terrestrial radiation (longwave), high level clouds are most
      important as they are coldest (the greenhouse effect depends on the
      temperature difference between the surface and the cloud). The end
      member case is 100\% coverage of high clouds which are optimised for
      their greenhouse effect, being both thicker and higher than our case
      study. Such an end member case gives a~forcing of 50\,{W\,m$^{-2}$},
      which would just be sufficient to resolve the FYSP. However, physical
      justification for any of the required changes is
      lacking. \citet{ron-lindzen-10} invoke a~controversial negative
      feedback of increased cirrus fraction with decreased temperature
      \citep[the \qut{iris} hypothesis of][]{lindzen-ea-01}, but a~true
      resolution to the FYSP should give temperatures equal or higher than
      present. Thus, even if the \qut{iris} hypothesis was correct, it would
      act to oppose warming. It is difficult to think of other mechanisms to
      make high clouds wider and thicker. Whether clouds should have been
      higher in the Archean may warrant more study. The absence of the
      strong stratospheric temperature inversion presently caused by ozone
      might contribute.  However, without increase in fraction or cloud
      water path, the forcing will likely be less than 5\,{W\,m$^{-2}$}.

      The question then naturally arises: How should one model early Earth
      climate? Some would look first towards a~general circulation model
      (GCM), in order to better represent the dynamics on which clouds
      depend. We disagree. Whilst dynamics are certainly important, it is
      unrealistic to think that in the near future clouds could be
      \textit{resolved} in a~global scale climate model applicable to
      palaeoclimate. Even in \qut{high resolution} models used for
      anthropogenic global change, cloud processes are parameterised
      sub-grid scale. As one moves towards deep palaeoclimate research, one
      moves further from the present atmospheric state for which the model
      may have been designed and can be validated. A~larger model therefore
      introduces greater, and harder to track, uncertainty. Considering what
      radiative forcing or warming a~given mixture of greenhouse gases will
      impart is a~first-order question, and one which should be answerable
      with a~first-order model. A~1-D model is sufficient for
      this, but clouds must be included. The appropriate starting point would likely be a model with fixed cloud optical depth and fixed cloud top temperature \citep[see, for example][]{reck-79,wang-stone-80}. For any proposed change to clouds, very great
      attention is needed to the feasibility of the mechanism involved. To
model these, one should probably look towards a~cloud microphysics resolving
model, coupled to appropriate models of CCN supply and chemistry.

A stronger greenhouse effect likely contributes the largest part of the forcing required to keep early Earth warm. It is important to remember, however, that the forcing from a greenhouse gas depends on the logarithm of its abundance. Thus, a modest forcing from clouds could have a large effect on how much of a greenhouse gas in needed; indicatively, a 10\,{W\,m$^{-2}$} from \chem{CO_2} requires a increase in concentration of a factor of 2 to 3. A large atmospheric \chem{CO_2} reservoir ($\sim 0.1$\,bar) may be slow to accumulate, as geochemical processes (principally volcanic outgassing) contributing linearly to concentration, so any other forcings may be rather useful in resolving the FYSP.

      In summary, it is necessary to include clouds in climate models if
      these are to be accurate. Resolution of the faint young sun paradox
      likely requires a~combination of a~few different warming
      mechanisms, including strong contributions from one or more greenhouse gases. Changed clouds could contribute warming, but this has yet to be
      justified -- and cooling caused by cloud changes is equally possible. Future work will no doubt propose novel mechanisms to change clouds. We hope that the results presented here will facilitate quick and accurate look-up the climatic effect of such changes. Proposed cloud-based resolutions with only limited
      greenhouse enhancement are not plausible.

\begin{acknowledgements}
Thanks to Richard Freedman for plotting the HITRAN data for
Fig.~\ref{f-plt_um_annotated} and William Rossow for providing the data for
Fig.~\ref{f-cloudfrac_4pcol}. Thanks to J.~Kasting and I.~Halevy for reviews and R. Pierrehumbert for comments. C.~G. was funded by a~NASA Postdoctoral Program
fellowship. K.~J.~Z was funded by the NASA Astrobiology Institute Ames Team
and the NASA Exobiology program.
\end{acknowledgements}


\end{document}